# HASPIDE-SPACE: a new concept of a radiation-resistant instrument for solar energetic particles

**C. Grimani[1,2], M. Fabi[1,2], M. Menichelli[3], F. Sabbatini[2] M. Villani[1], S. Aziz[4,5], L. Barraud[6], A. Bashiri[7,8], C. Buti[2,9], L. Calcagnile[4,5], D. Calvo[10], D. Caputo[11,12], A. P. Caricato[4,5], R. Catalano[13], M. Cazzanelli[14], R. Cirio[10], G. A. P. Cirrone[13], F. Cittadini[3,15], T. Croci[3,15], G. Cuttone[13], G. de Cesare[11,12], P. De Remigis[10], S. Dunand[6], L. Frontini[16,17], M. Guarrera[13], H. Hasnaoui[14], G. Insero[2,9], M. Ionica[9], K. Kanxheri[3,18], G. Konstantinou[6], M. Large[7], F. Lenta[10,19], V. Liberali[16,17], N. Lovecchio[11,12], M. Martino[4,5], G. Maruccio[4,5], G. Mazza[10], A. G. Monteduro[4,5], A. Morozzi[3], A. Nascetti[11,20], S. Pallotta[2,9], D. Passeri[3,21], M. Pedio[22,3], M. Petasecca[7], G. Petringa[13], F. Peverini[3,18], P. Placidi[3,21], M. Polo[14], A. Quaranta[14], G. Quarta[4,5], S. Rizzato[4,5], A. Stabile[16,17], C. Talamonti[2,9], L. Tosti[3], M. M. Vasquez[16], R. J. Wheadon[10], N. Wyrsch[6], N. Zema[3,23], L. Servoli[3]**

[1] Dipartimento di Scienze Pure e Applicate (DiSPeA), Università di Urbino Carlo Bo, Via S. Chiara 27, 61029 Urbino, Italy

[2] Istituto Nazionale di Fisica Nucleare (INFN), Sezione di Firenze, Via B. Rossi 1, 50019 Sesto Fiorentino, Firenze, Italy

[3] Istituto Nazionale di Fisica Nucleare (INFN), Sezione di Perugia, Via A. Pascoli snc, 06123 Perugia, Italy

[4] Istituto Nazionale di Fisica Nucleare (INFN), Sezione di Lecce, Via per Arnesano, 73100 Lecce, Italy

[5] Dipartimento di Matematica e Fisica “Ennio de Giorgi” Università del Salento, Via per Arnesano, 73100 Lecce, Italy

[6] École Polytechnique Fédérale de Lausanne (EPFL), Photovoltaic and Thin-Film Electronics Laboratory (PV-Lab), Rue de la Maladière 71b, 2000 Neuchâtel, Switzerland

[7] Centre for Medical Radiation Physics, University of Wollongong, Northfields Ave Wollongong NSW 2522, Australia

[8] Najran University, Faculty of Arts and Sciences, Department of Physics, King Abdulaziz Rd, 1988 Najran, Saudi Arabia

[9] Dipartimento di Scienze Biomediche sperimentali e Cliniche “Mario Serio”, Università di Firenze, Viale Morgagni 50, 50134, Firenze, FI, Italy

[10] Istituto Nazionale di Fisica Nucleare (INFN), Sezione di Torino, Via Pietro Giuria 1, 10125 Torino, Italy

[11] Istituto Nazionale di Fisica Nucleare (INFN), Sezione di Roma 1, Piazzale Aldo

Moro 2, Roma, Italy
[12] Dipartimento Ingegneria dell'Informazione, Elettronica e Telecomunicazioni, dell'Università degli studi di Roma via Eudossiana, 18 00184 Roma, Italy
[13] Istituto Nazionale di Fisica Nucleare (INFN), Laboratori Nazionali del Sud, Via S. Sofia 62, 95123 Catania, Italy
[14] Istituto Nazionale di Fisica Nucleare (INFN), TIFPA c/o Dipartimento di fisica dell'Università di Trento, Via Sommarive 14, 38123 Povo, Trento, Italy
[15] Dipartimento di Fisica e Astronomia Università di Padova, via Marzolo 8, 35131 Padova, Italy
[16] Istituto Nazionale di Fisica Nucleare (INFN), Sezione di Milano, Via Celoria 16, 20133 Milano, Italy
[17] Dipartimento di Fisica dell'Università degli Studi di Milano, via Celoria 16, 20133 Milano, Italy
[18] Università di Perugia, Dipartimento di Fisica e Geologia, Via A. Pascoli snc, 06123 Perugia, Italy
[19] Politecnico di Torino, Corso Duca degli Abruzzi 24, 10129 Torino, Italy
[20] Scuola di Ingegneria Aerospaziale Università degli studi di Roma. Via Salaria 851/881, 00138 Roma. Italy
[21] Università di Perugia, Dipartimento di Ingegneria, Via Goffredo Duranti 93, 06125 Perugia, Italy
[22] Istituto Officina dei Materiali CNR, Via A. Pascoli snc, 06123 Perugia, Italy
[23] CNR Istituto struttura della Materia, Via Fosso del Cavaliere 100, Roma, Italy

E-mail: catia.grimani@uniurb.it

**Abstract.** Galactic cosmic rays and solar energetic particles (SEPs) affect the performance of instruments carried on board space missions and are the source of the absorbed dose to astronauts. Particles above 100 MeV are the most penetrating. The overall particle flux impacting spacecraft increases by several orders of magnitude during the most intense SEP events and gamma-ray bursts, compared to the galactic cosmic-ray background. Unfortunately, most detectors dedicated to the monitoring of solar particles in space do not allow measurements above 200 MeV. Space agencies and the space and heliospheric weather communities are interested in improving this scenario. We propose a new, compact, low-power, and radiation-resistant instrument, intended for long-duration missions to monitor solar X-rays, electrons, and protons up to energies of tens of keV, MeV, and 600 MeV, respectively. Intense long and short gamma-ray bursts would be observed in the absence of solar activity.

## 1. Introduction

Particles with energies above 100 MeV penetrate more than 10 g cm$^{-2}$ of spacecraft material, thus limiting the performance of on-board instrumentation and constituting the origin of the absorbed doses on board manned and unmanned space missions [1, 2, 3, 4, 5]. High-energy radiation in the interplanetary medium consists mainly of charged particles of galactic and solar origin. Protons and nuclei make up the majority of galactic cosmic rays (GCRs) which present energies ranging between tens of MeV/n and $3\times10^{20}$ eV [6]. The flux of GCRs undergoes long-term variations ($>$ 1 month) associated with the quasi-eleven-year solar cycle (Schwabe cycle) and the approximately twenty-two-year global solar magnetic field (GSMF) polarity reversal (Hale cycle; [7, 8]). Galactic cosmic rays also show short-term variations ($<$ 1 month) due to the passage of large-scale interplanetary structures such as high-speed solar wind streams and interplanetary counterparts of coronal mass ejections (CMEs; [9, 10, 11, 12]). Long- [13, 14, 15, 16] and short-term [17] modulations account for variations of the overall flux of GCRs of up to a factor of 4 and 50% at 1 AU, respectively. The flux of particles of solar origin may exceed the GCR background by several orders of magnitude up to ten days per month at solar maximum [16]. Solar energetic particle (SEP) events are divided into two classes: impulsive and gradual [18]. The first class appears electron rich and presents proton acceleration typically below 50 MeV, while the second class is characterized by protons accelerated at shocks driven by CMEs. Gradual events are the main focus of our research, as they have the greatest impact on manned and unmanned space missions. Meter-wavelength type II solar radio bursts, associated with shock-accelerated electrons in the solar corona [19], represent the signature and the precursor of high-energy proton detection. X-ray emission from solar flares often accompanies the initial phases of particle acceleration [15, 16, 17].

Gradual SEP events occur randomly, although the annual fluence of solar particles can be estimated within an uncertainty of a factor of two based on observations of the yearly sunspot number [20, 21]. In a previous work [22], we have emphasized that particle detectors hosted on board most missions dedicated to solar and heliospheric physics allow the measurement of solar particles up to 200-250 MeV. This scenario can possibly be ascribed to the evidence that solar particle fluxes exhibit spectral indices ranging from 3 to 4 above tens of MeV. Therefore, in most cases, solar particle fluxes are poorly populated at high energies. Nevertheless, this is not the case during the most intense events when high internal charging and absorbed doses on board satellites are observed along with instrument malfunctions [3, 4].

Papaioannou et al. [23] made an attempt to reconcile measurements of low-energy solar particles in space and neutron monitor data during ground level enhancements. Unfortunately, normalization problems between ground-based and space-based data, as well as different observation times, limit the quality of the results. Similar problems are encountered when multispacecraft observations of the same events are compared to models of solar particle propagation in the interplanetary medium. The AMS-02 Collaboration has recently presented a preliminary work on near-Earth SEP detection both at low and high energies [24, 25]. AMS-02 is a magnetic spectrometer experiment aboard the Space Station that allows precise measurements of the GCR flux spectra only above hundreds of MeV due to the high average geomagnetic cut-off along its entire orbit. Despite this, AMS-02 was able to monitor high SEP fluxes with a small statistical uncertainty down to a few tens of MeV for less than one hour per day, when the space station orbited in regions of low geomagnetic cut-off, providing a discontinuous observation of the evolution of the events.

A proposal for the detection of solar energetic particles up to hundreds of MeV has also been made recently by O'Neill et al. [26] with the HEPI instrument based on the detection of protons through the emission of Cherenkov radiation and the use of silicon photomultipliers (SiPM). Furthermore, continuous monitoring of solar proton and helium differential fluxes from 70 MeV/n through 400 MeV/n will be performed on board the future gravitational wave laser interferometer in space, LISA [27, 28, 29], after the launch of the mission in 2035. Both LISA radiation monitors and HEPI are based on SiPM technology characterized by the same radiation hardness of silicon, typically $10^{14}$ protons cm$^{-2}$ [30]. Multipoint observations of SEP events with low-cost, lightweight, low-power-consumption detectors hosted aboard long duration space missions may require the use of instruments with better radiation hardness than crystalline silicon [31]. Valuable data on the dynamics of rare and intense SEP events in terms of the propagation of solar particles in the interplanetary medium and the distribution of particle pitch angles are absolutely necessary. We recall that the SEP pitch angle is defined as the angle between the velocity of particles and the nominal direction of the interplanetary magnetic field vector.

The instrument we are proposing, based on hydrogenated amorphous silicon sensors (a-Si:H), has a higher radiation hardness than crystalline silicon and can provide very precious insights into the acceleration and propagation of SEPs up to hundreds of MeV. Our instrument can also provide valuable clues into high-energy astrophysics in case of rapid increase in the fluxes of X rays and $\gamma$ rays not associated with solar flares but with $\gamma$-ray bursts [see for details 32]. As a matter of fact, the a-Si:H substrate is resistant to total ionizing doses of up to 100 Mrad [33], displacement damage up to a fluence of $10^{16}$ 1 MeV neutron equivalent cm$^{-2}$ [34] and combined damage up to a fluence of $10^{16}$ 24 GeV protons cm$^{-2}$ [35].

Our prototype instrument was developed in the framework of the Work Package 5 of the HASPIDE (Hydrogenated Amorphous Silicon PIxel DEtectors for ionizing radiation) project funded by the Italian Institute for Nuclear Physics. The sensitive layers of the instrument consist of matrices of a-Si:H sensors. These devices were fabricated via the standard PECVD (Plasma Enhanced Chemical Vapor Deposition), with thickness ranging from hundreds of nanometers to ten micrometers. Low-temperature deposition is made on substrates of plastic materials like Polyimide. The resulting devices are mechanically flexible, very thin and lightweight, adaptable to curved configurations, such as cylinders, to assure high-coverage solid angles [36, 37, 38, 39]. The performance of single sensors meant for medical use were tested with 3 MeV protons, 6 MeV electrons and 6 MV photon beams [see 22, and references therein]. FLUKA Monte Carlo simulations [40, 41, 42] allowed us to estimate the optimal size and thickness of the sensors to maximize the sensitivity of the instrument for space applications. At the time of writing, our prototype instrument is being assembled for a test on a proton beam in the tens to hundreds of MeV range scheduled for the next months at TIFPA in Trento (Italy). In this work, we discuss the expected performance of the prototype during SEP events.

## 2. HASPIDE-SPACE instrument design

In [22] we have demonstrated that hydrogenated amorphous silicon sensors developed for medical applications could be operated with electric fields of 1.5-4.0 V/$\mu$m or no bias at all. In principle, this latter feature would be very important in space applications for which power consumption must be kept as low as possible for the design of active dosimeters and instruments to be carried on board space missions. Unfortunately, a sensitivity decrease of our devices is observed with no bias applied. In [22] we have also shown that soft and hard X rays associated with X and M flares [43], typically observed during medium-strong solar activity, were detectable with a-Si:H sensors used for medical applications. Unfortunately, this was not the case with protons and electrons except for extreme gradual and impulsive events, respectively. This scenario motivated our effort to design an instrument that consists of sensitive and passive layers for monitoring medium-to-strong events of solar protons ($> 10^7$ protons/cm$^{-2}$). A series of Monte Carlo simulations were used to optimize the design of our prototype instrument for space applications aimed essentially at monitoring space weather. In particular, we estimated the minimum area and thickness of a-Si:H sensors that would have allowed us to detect SEP proton fluxes from a few MeV up to at least 600 MeV with proper arrangement of active and passive material layers. These requirements were met with four sensitive planes of 3x3 a-Si:H sensor arrays of 7x7 mm$^2$ active area and 10 $\mu$m thickness (see Figure 1) and three layers of tungsten for a total thickness of 8 cm. The sensors, sketched in Figure 2, have an hybrid structure. On top of the Polyimide substrate there is a layer deposition of Aluminum and Chrome. Above this deposition there is an n-doped a-Si:H layer. On top of this, a thick layer (10 µm) of intrinsic a-Si:H is deposited. Finally, there is a square patterned layer of MoOx covered with an Indium Tin Oxide (ITO) conductive layer to make the top electric contact. This device is called hybrid because it has a combination of a n-i-p device (n-i junction) on the bottom part, and a charge selective contact device (i-a-Si:H/MoOx junction) on top.

The a-Si:H devices are produced by colleagues at EPFL (École Polytechnique Fédérale de Lausanne - Neuchâtel, Switzerland). In Figure 1 it is possible to see the bond for each individual device anode (green squares), while the cathode is common to the nine devices of each sensitive plane and is connected on the metal chromium layer indicated by the red arrow in the top-right part of the picture. We balanced the sensitivity and power consumption of these sensors with bias electric field of 1 V/$\mu$m, for a total of 10 V applied voltage. By comparing Figure 3 and figure 9 in [22], it can be noticed that the current noise of the new sensors is about 40% smaller than that of sensors used for medical applications. These results represent a significant improvement in the sensitivity of the sensors, allowing the detection of protons with a signal-to-noise ratio of 5 with an energy deposit of 60 MeV/s compared to the 105 MeV/s discussed in [22]. The characterization of these sensors shows a linear response of the detector in current with dose rate, within an uncertainty of 7% as shown in Figure 4.

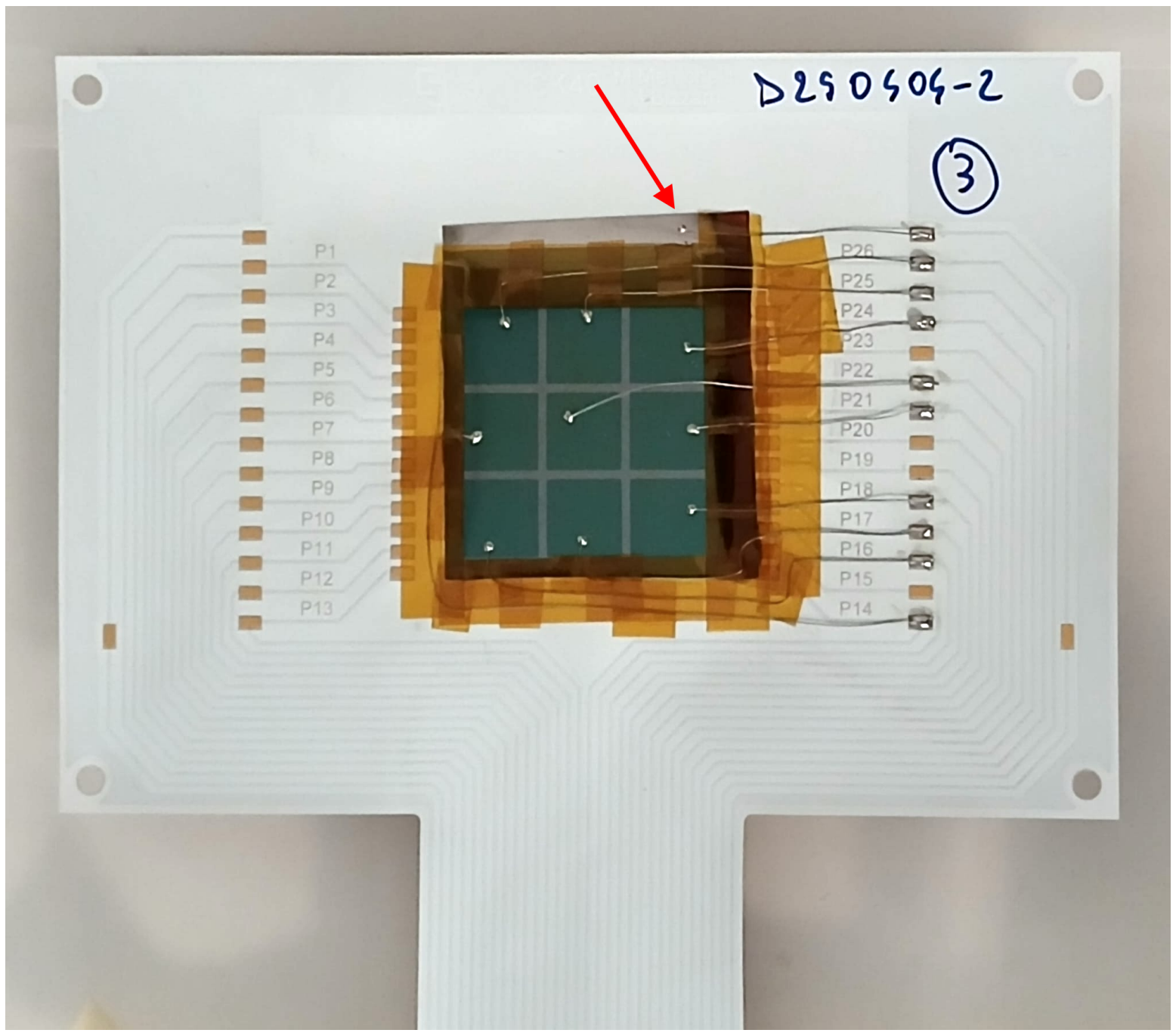

Figure 1: Active layer of the HASPIDE-SPACE instrument prototype (see text for details).

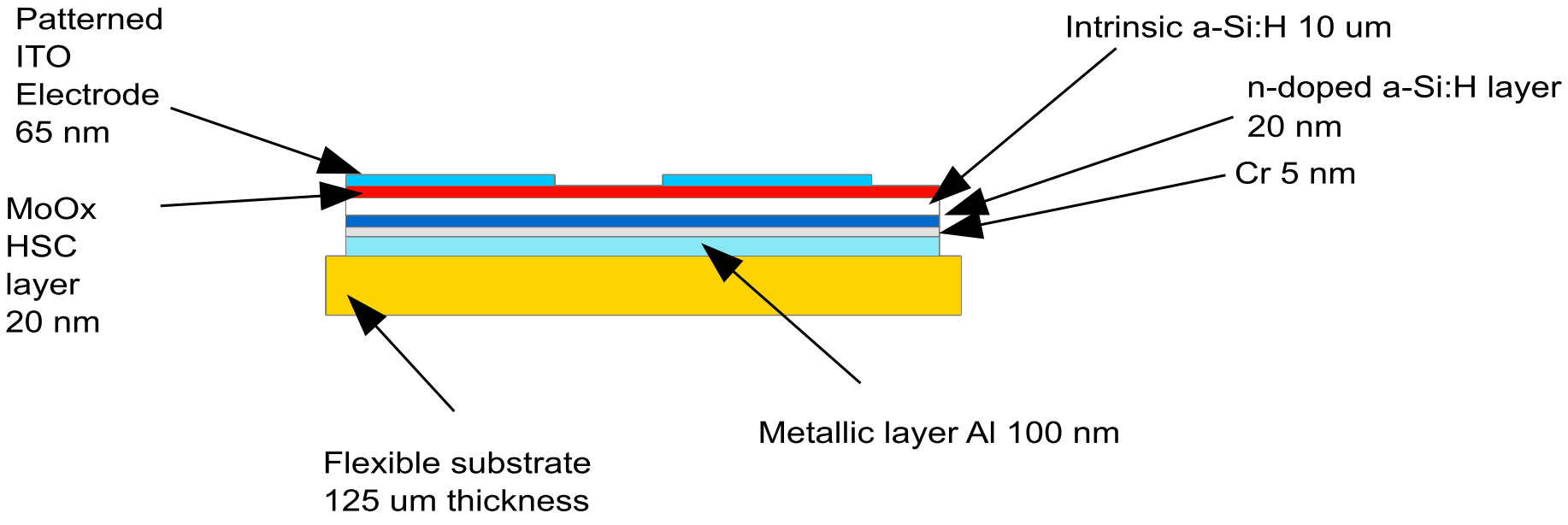

Figure 2: Sketch of the structure of a-Si:H devices developed for HASPIDE-SPACE.

In Figure 5 (left image) we show the combinatorial geometry of the prototype instrument implemented in the Monte Carlo simulations using Flair‡ [40] for FLUKA. Four sensitive layers of device arrays (gold thin layers A, B, C, D in the right part of Figure 5) are placed among three passive layers of tungsten of 2 mm, 1 cm and 6.8 cm thickness and $2.4 \times 2.4$ cm$^2$ area (blue blocks in Figure 5). The height of the instrument, including the mechanical supports, is 12 cm and the total mass amounts to 1.3 kg. The orange side sensors (F1, F2, F3, F4 and G1, G2, G3 and G4) deposited on yellow kapton support in the right image of Figure 5 are placed on each side of the tower at the top and bottom of the thicker tungsten block. The side sensors are necessary in space to estimate the direction of the incoming solar particles and the amount of particles escaping the instrument. The CAD§ model of the mechanical supports designed for both active and passive planes for the beam test of the prototype instrument are depicted in Figure 6. On the left side of the figure, we show the support used to hold the kapton layers with the active sensor arrays visible in Figure 1. The right support contains a housing to hold the tungsten absorbers. The thicker tungsten blocks are held at both ends.

‡ Flair (`https://flair.web.cern.ch/flair/`) is a Python-based graphical front-end interface of the FLUKA Monte Carlo program. A supplemental Flair module called geoedit (geometry editor) allows to build and debug complex combinatorial geometries

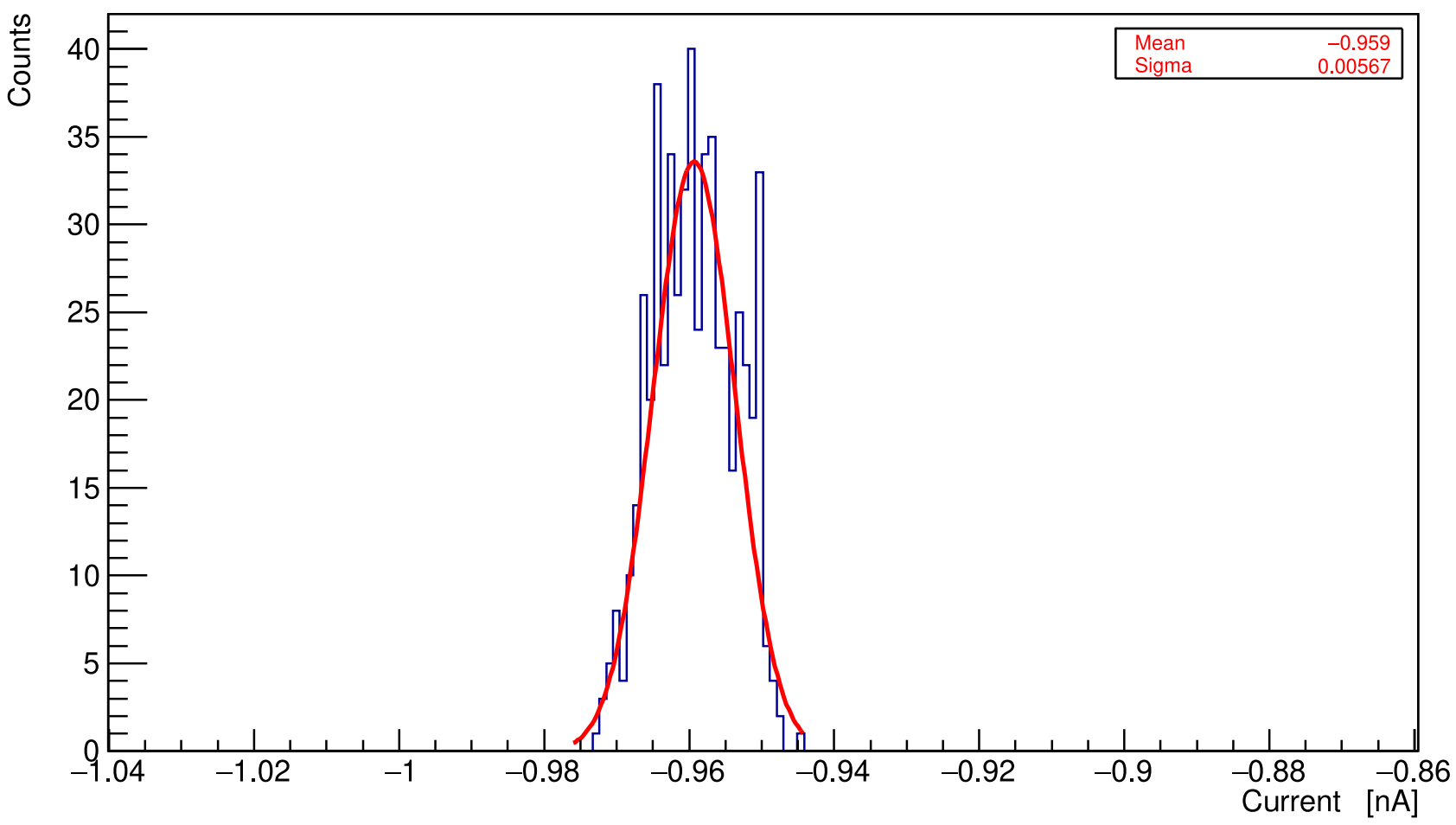

Figure 3: Sensor current noise represented by the width of the gaussian peak.

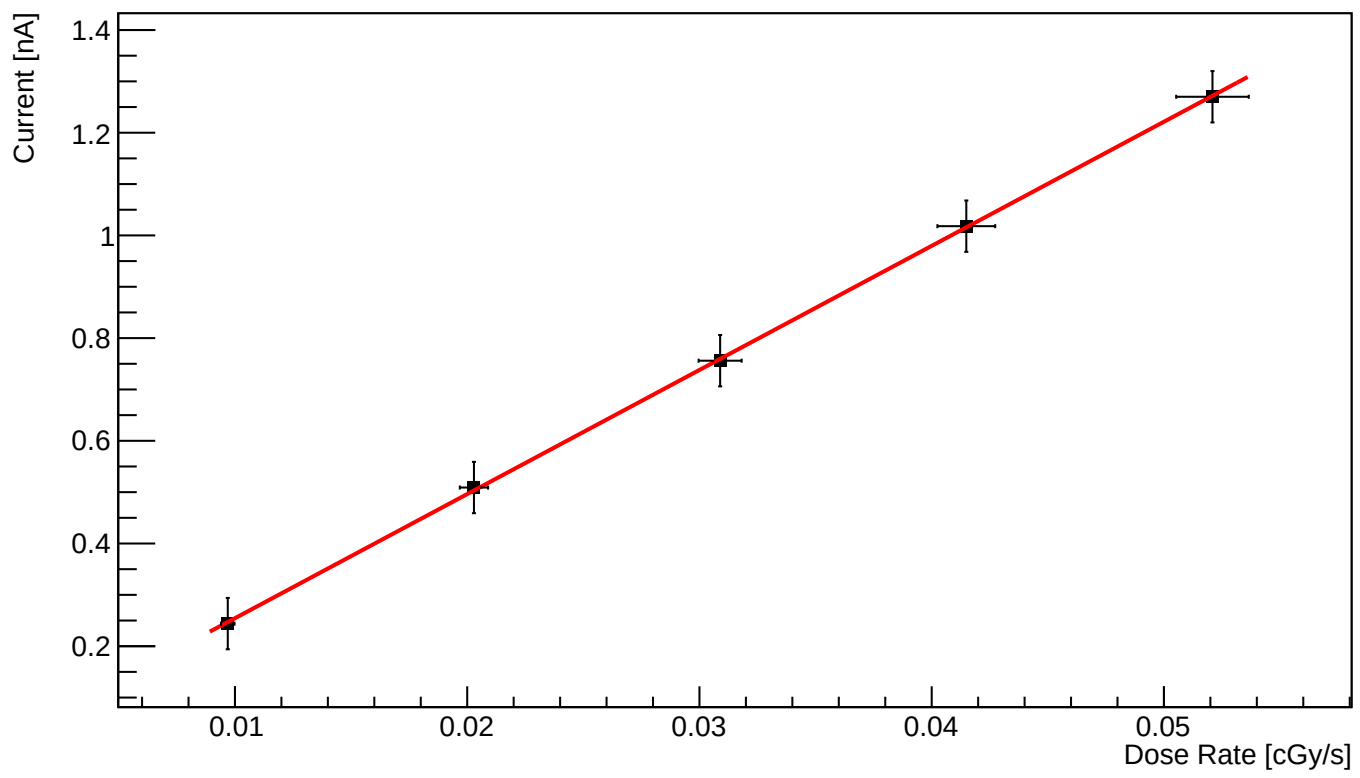

Figure 4: HASPIDE-SPACE sensor linear response in current with dose rate.

For the readout of these a-Si:H sensors, a 12-channel custom ASIC prototype

§ See `https://www.freecad.org/`

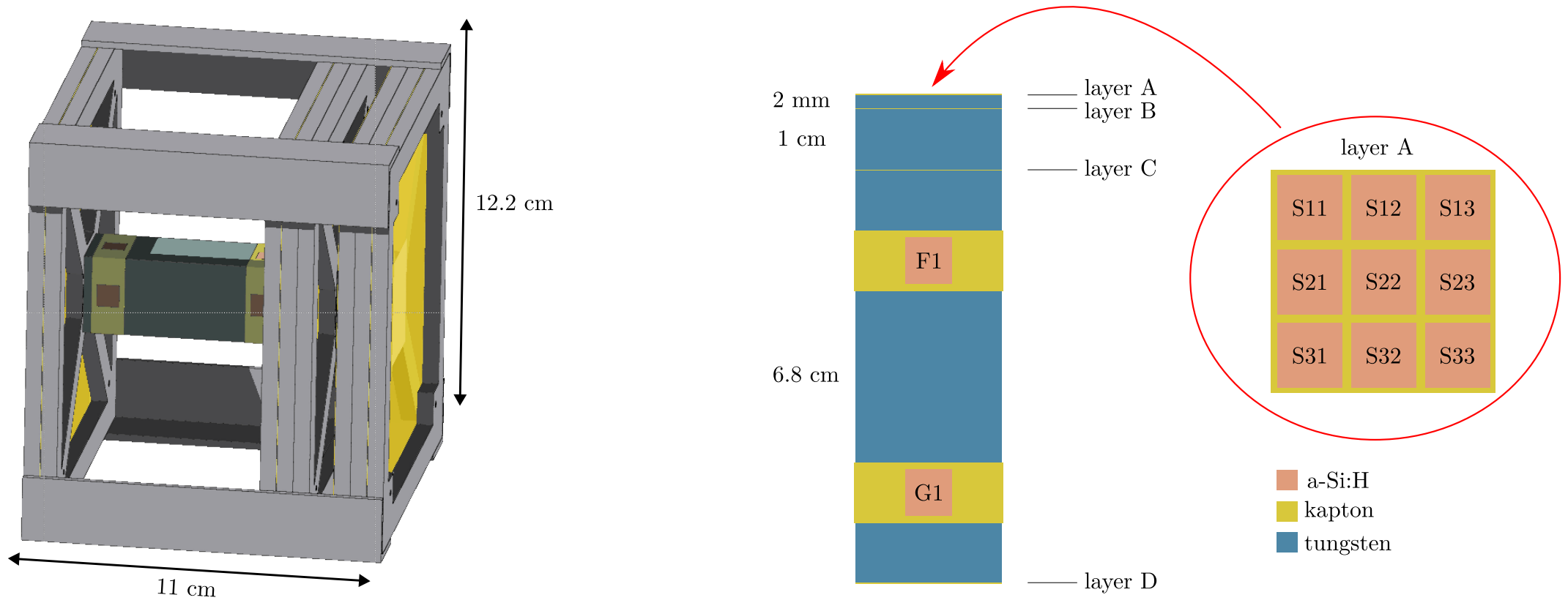

Figure 5: Left: Monte Carlo combinatorial geometry of the HASPIDE-SPACE prototype instrument. Active (gold) and passive (blue) layers are arranged in a tower-shape instrument. Four L-shaped profiles are used to keep in place the instrument. Right: side and top views of the detector arrangement. The yellow thin layers A, B, C and D constitute the sensitive planes made of matrices of a-Si:H sensors. The blue regions represent the tungsten blocks. Side sensors (F1 and G1) are shown in orange. The other side sensors F2-F4 and G2-G4 are placed on the sides of the tower not shown in the figure.

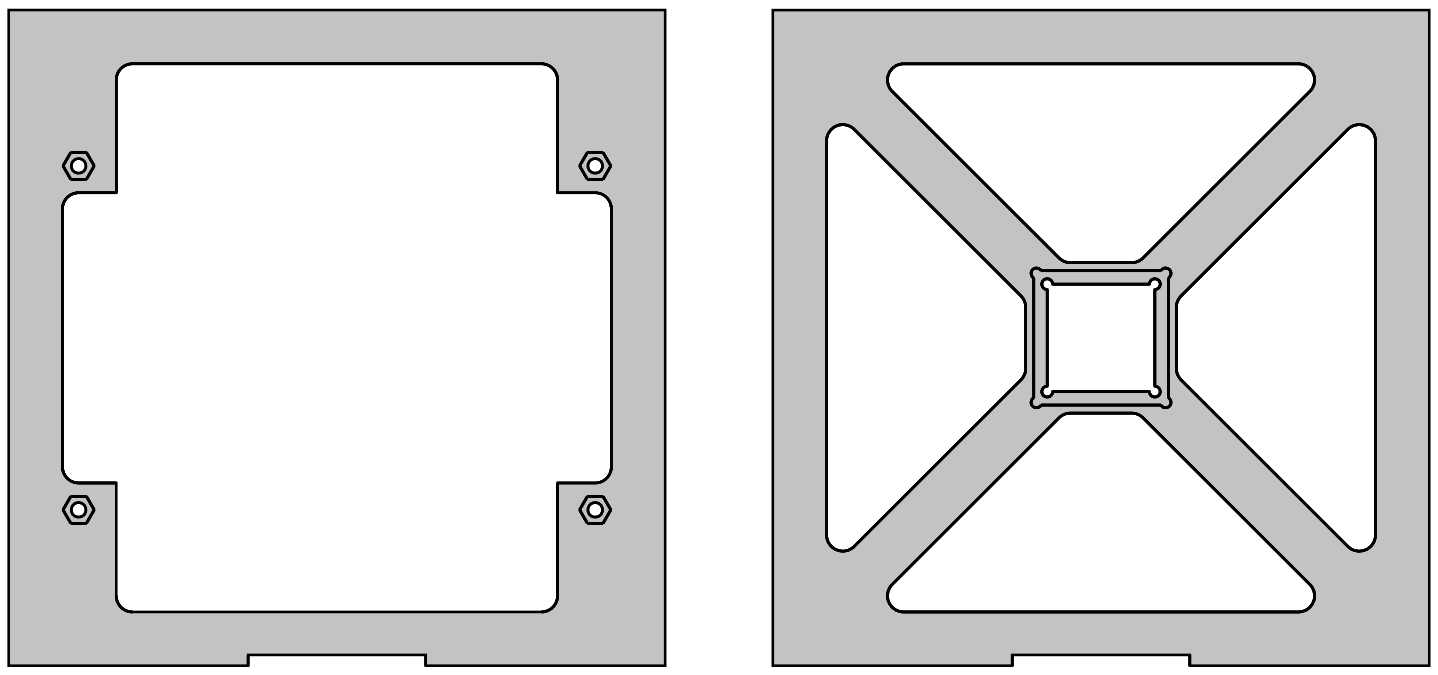

Figure 6: Left panel: Mechanical support of active arrays of sensors. Right panel: Mechanical support of the tungsten blocks that play the role of the passive parts of the instrument.

(named Cleopatra) has been developed using a commercial CMOS 28 nm technology. This readout chip is based on the recycling integration principle [44, 45] and provides current-to-frequency conversion achieving large dynamic range and direct digital output. Each analog input channel includes a Charge Sensitive Amplifier (CSA), a discriminator with a programmable threshold, a charge injection circuit and a test-pulse control logic. During the circuit operation, the feedback capacitor $C_f$ of the CSA is continuously charged by the input current $I_{in}$ and its output voltage $V_{CSA}$ increases. The slew rate of the output voltage of the CSA is proportional to the input current: $\frac{dV_{CSA}}{dt} = \frac{I_{in}}{C_f}$. The voltage output of the CSA is sent to the input of a discriminator. When the ramp crosses the discriminator threshold, a fixed amount of charge Q is subtracted from the amplifier input, reducing $V_{CSA}$ by the amount $\Delta V=\frac{Q}{C_f}$. The measurement of the input current can therefore be obtained by the product of the number of pulses generated during the time window multiplied by the amount of charge subtracted at each pulse according to the formula:

$$I_{in} = f_{OUT}\, Q, \tag{1}$$

where $f_{OUT}$ is the frequency of the injection signal.

The readout chip has been tested by injecting a precise voltage with a Source Measurement Unit (SMU) and measuring the current with the same instrument using a series 10 MΩ resistor in order to have a high impedance source. The digital interface is controlled by an FPGA-based evaluation board connected to a computer via an Ethernet link. The measured non-linearity is below 0.5% for currents above a few nA. In a wider input current range (using a 1 MΩ series resistor) up to 3.5 $\mu$A a non-linearity below 2% was measured.

## 3. Solar activity and solar energetic particle event occurrence

The Sun shows a long-term activity following a quasi 11-year cycle (Schwabe cycle) and shows the same polarity of its GSMF approximately every 22 years (Hale cycle; [16, and references therein]). It is worthwhile to recall that the polarity of the GSMF is defined positive(negative) when the magnetic field lines of force exit from(enter into) the Sun North Pole. The solar activity and the GSMF modulate the flux of GCRs propagating from the interstellar medium to the inner heliosphere. Galactic cosmic rays consist of 90% protons, 8% helium nuclei, 1% heavy nuclei and 1% electrons. The minimum and maximum GCR proton flux observed during the last three solar cycles are shown in Figure 7 [46].

We emphasize that extreme SEP events can be observed at both solar minimum and solar maximum, although, on average, a larger number of SEP events occur at solar maximum. The intensity of the present solar cycle 25 is consistent with average solar cycles over the last hundred years. The same is expected for solar cycle 26 [16].

As recalled above, SEP events are divided in two main categories: impulsive and gradual [47, and references therein]. Gradual SEP events consist of 99% of protons and 1% electrons shock-accelerated by propagating CMEs. These events are at the origin of the dramatic increase of the internal charging and dose absorption in satellites. Therefore, in the following, we focus in particular on SEP events characterized by particle acceleration above tens of MeV. In Figure 8 we report the flux of protons observed at the onset and peak of a sample of SEP events of different intensity [17, 48]. Comparison of this figure with Figure 7 reveals that the overall flux of protons incident on a spacecraft may increase by several orders of magnitude during SEP events. The detection of proton flux enhancements typically occurs within tens of minutes after the observation of solar eruptions at the origin of particle acceleration when the spacecraft is well magnetically connected to the associated active regions. As a matter of fact, the time profiles of SEPs depend on the evolution of the intersection point between the magnetic field line connecting the spacecraft to the shock front, produced by the propagating interplanetary counterparts of CMEs, where the particles are accelerated. Consequently, the spatial distribution of solar particles changes during the evolution of events and tends to become isotropic during the decay phase. The determination of the spatial distribution of solar particles must be correctly taken into account to estimate the instrument geometrical factor and the SEP flux from particle count. To this purpose, in [49] we estimated the varying geometrical acceptance of the particle detector on board the LISA Pathfinder mission during SEP events. In order to study the spatial distribution of SEPs with HASPIDE-SPACE we will benefit from the instrument side sensors and Monte Carlo simulations. We will further optimize the detector design relative to the current prototype, based on the performance observed in the proton beam test up to hundreds of MeV. Our final aim is to fly our instrument during the next solar maximum.

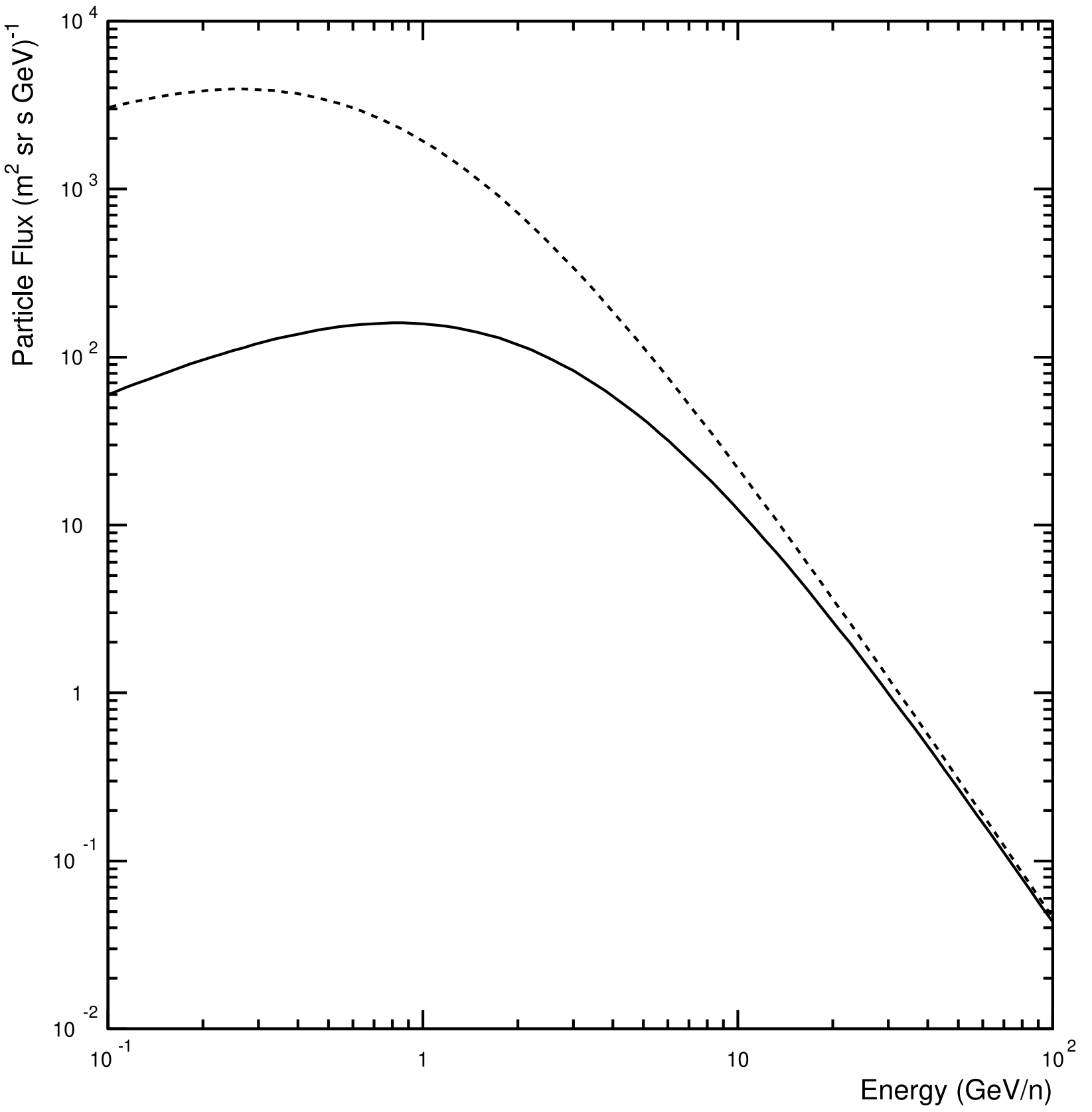

Figure 7: Minimum (bottom continuous line) and maximum (top dashed line) GCR proton observations during the last three solar cycles.

We have carried out a study on the occurrence of SEP events during solar cycle

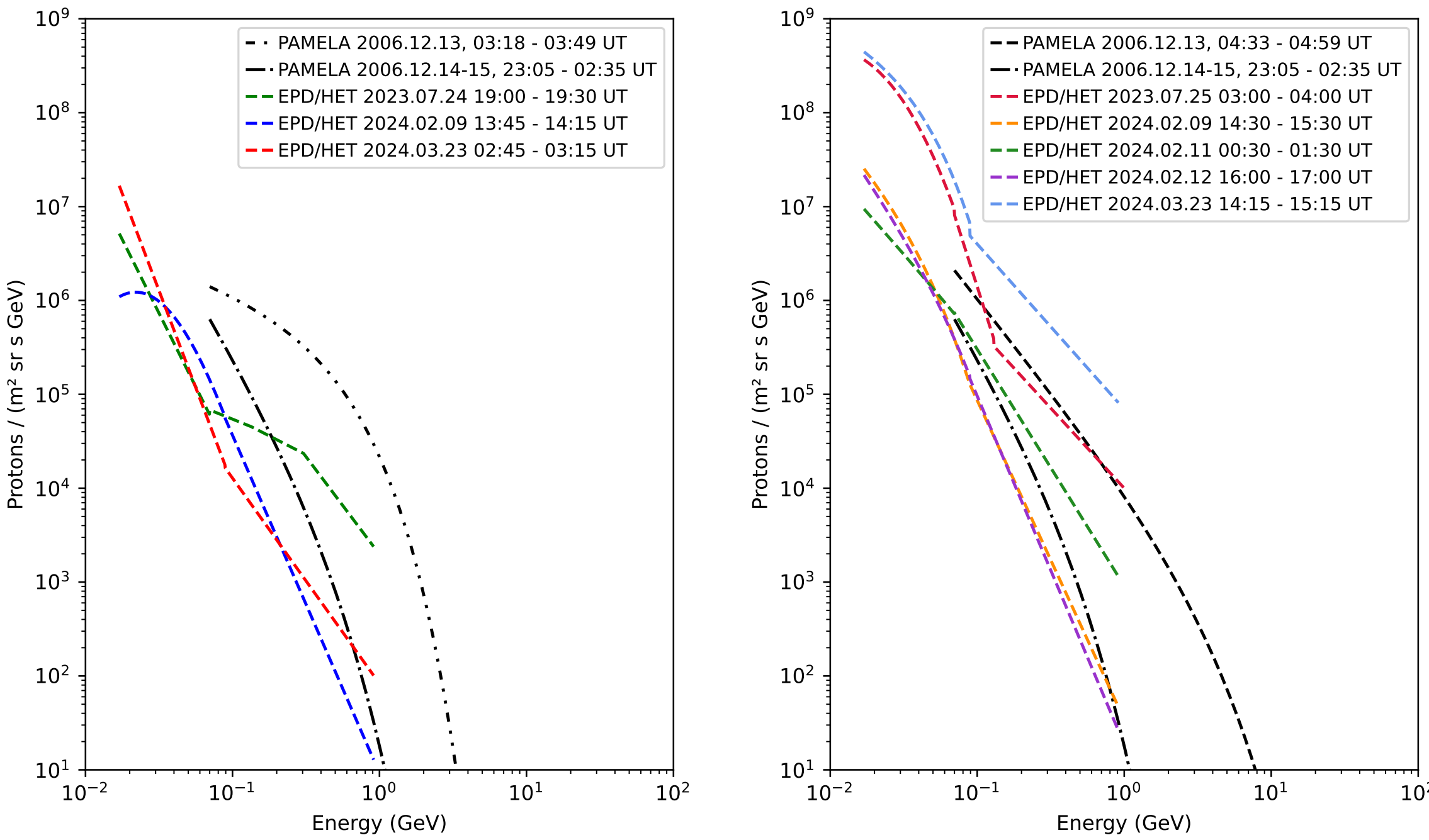

Figure 8: Left panel. Proton flux at the onset of various SEP events measured by the near-Earth satellite PAMELA experiment and the EPD/HET experiment on board Solar Orbiter. Right panel. Same as the left panel at the peaks of the events.

26 for the LISA space interferometer. In that work [16], we estimated the minimum, average and maximum number of events with a fluence above $10^6$ protons cm$^{-2}$ with the Nymmik model [20, 21]. According to these estimates, the maximum number of SEP events that could be observed with the HASPIDE-SPACE during the next solar maximum would be of the order of 10 per year in the period 2035-2037.

The ability to observe gamma-ray bursts in the absence of M and X flares would also would also provide valuable contributions to high-energy astrophysics and space weather science.

## 4. A case study: the solar energetic particle event of March 23, 2024

In this section, we discuss the evolution of the activity on the Sun at the origin of the SEP event dated March 23, 2024 as a case study for HASPIDE-SPACE. This event was fully monitored with the Solar Orbiter observatory ‖ ¶ and near-Earth instruments. The proton flux observation profile is shown in Figure 9. The onset of the event occurred after 02:00 UT at different times depending on particle energy (see Figure 10). The overall particle flux increased by almost five orders of magnitude at 10-20 MeV (red curve) and about three in the energy range 80-90 MeV (light blue curve). The peak of this event in this last energy range was observed with Solar Orbiter between 14:15 and 15:15 UT. The proton differential flux observed during selected intervals of time representing the whole dynamics of the event is reported in Figure 10. During the SEP event Solar Orbiter was approximately radially aligned with Earth (+8° in longitude) and only 0.39 AU from the Sun, as shown in Figure 11. Proton observations made near Earth by GOES-18 in similar energy ranges[+] are shown in Figure 9 at the onset and at the peak of the event for comparison with Solar Orbiter/EPD data [48]. It is interesting to note, in the inset of the figure, the different times of the onset of the event due to the different radial distances of the two spacecraft from the Sun. Unfortunately, the GOES-18 data are affected by large fluctuations. Nevertheless, multispacecraft observations are of primary importance for SEP propagation models and space weather science investigations.

‖ Data up to 100 MeV are publicly available at `https://soar.esac.esa.int/soar/`

¶ Best fit to the data above 100 MeV, kindly provided by the EPD/HET Collaboration, were presented at SPACEMON 2025 `https://indico.esa.int/event/555/contributions/10791/`

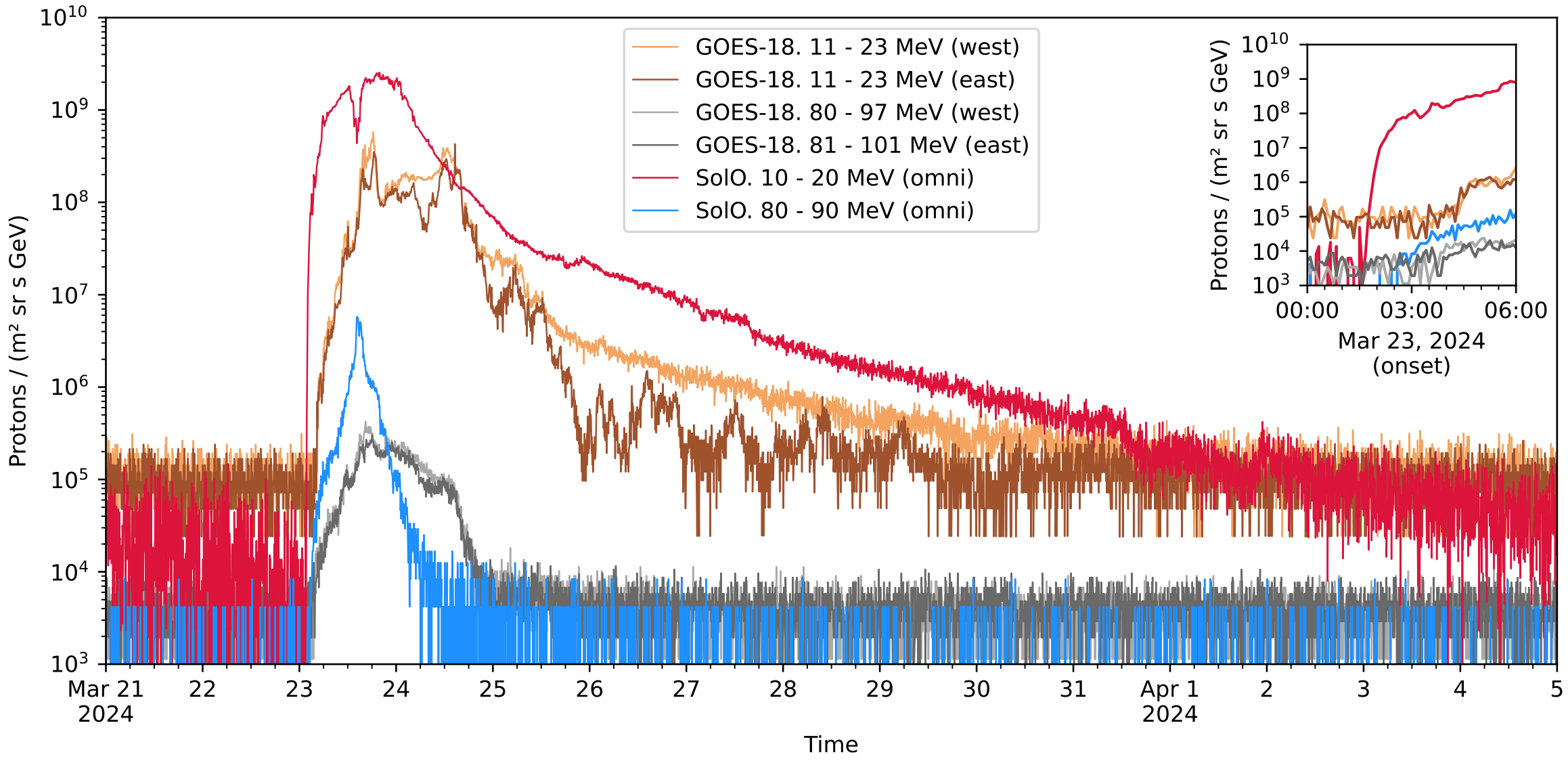

Figure 9: Proton differential flux profiles above the GCR background (blue) during the March 23, 2024 SEP event observed with the EPD/HET instrument on board Solar Orbiter and with GOES-18 near Earth.

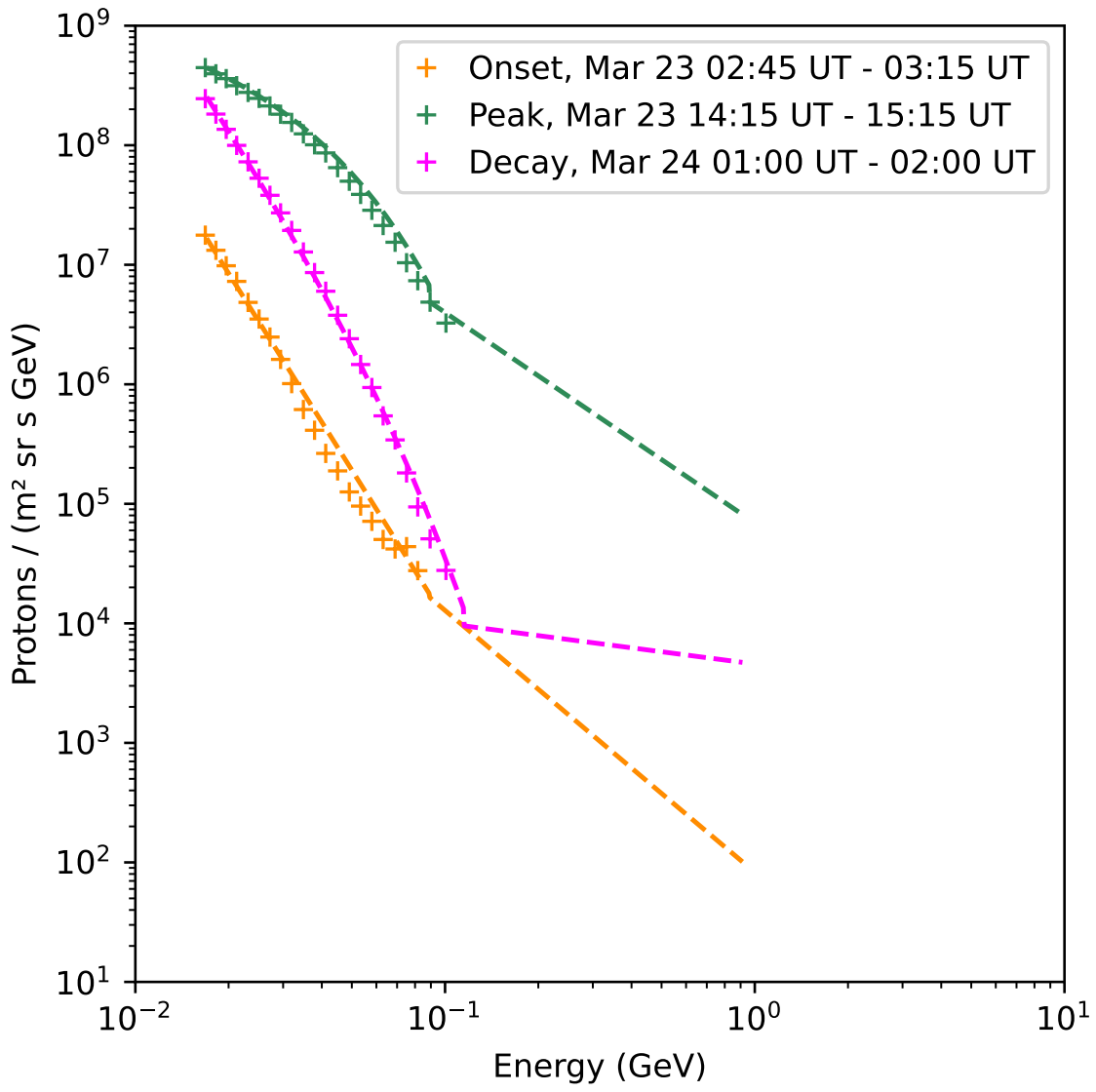

Figure 10: Proton energy differential flux evolution measured by the EPD/HET instrument aboard Solar Orbiter during the SEP event dated March 23, 2024.

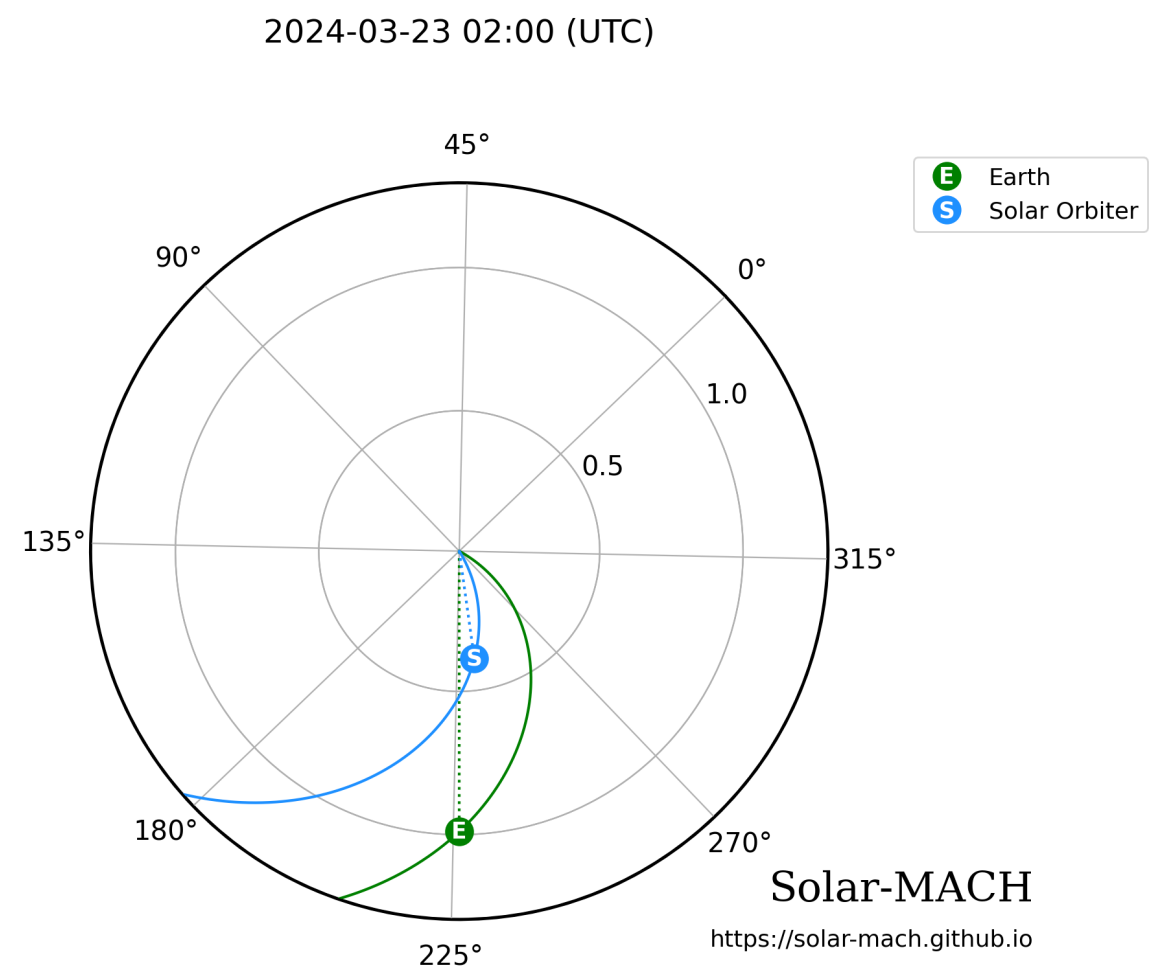

Figure 11: View of the ecliptic plane from solar north on March 23, 2024 at 02:00 UT. The interplanetary magnetic field Parker spirals are shown at the positions of Solar Orbiter (S) and Earth (E). Solar Orbiter was at 0.39 AU from the Sun and at +8° in longitude compared to Earth. The image was obtained with the Solar MAgnetic Connection Haus tool [50].

### *4.1. Solar activity, X-ray, and electron fluxes associated with the event*

An X1.12 flare, most likely associated with the active region 3614 located in the North-East quadrant of the Sun (N27E08; see Figure 12), was observed to start at 00:58 UT, peak at 01:33 UT and end at 02:22 UT by GOES-18 near Earth (see Figure 13). The X-ray emission appearing in Figure 14 observed by STIX [51]♯ aboard Solar Orbiter began at 00:52 UT due to the different radial distances of the two spacecraft from the Sun. Unfortunately, an attenuator was used for the STIX instrument, which prevented the measurement of the absolute differential photon flux between 4 and 10 keV beyond $10^6$ counts in 4-second binned time intervals. A halo CME with speed of 1470 km $s^{-1}$

[+] Data gathered from `https://data.ngdc.noaa.gov/platforms/solar-space-observing-satellites/goes/goes18/l2/data`
♯ `https://datacenter.stix.i4ds.net/`

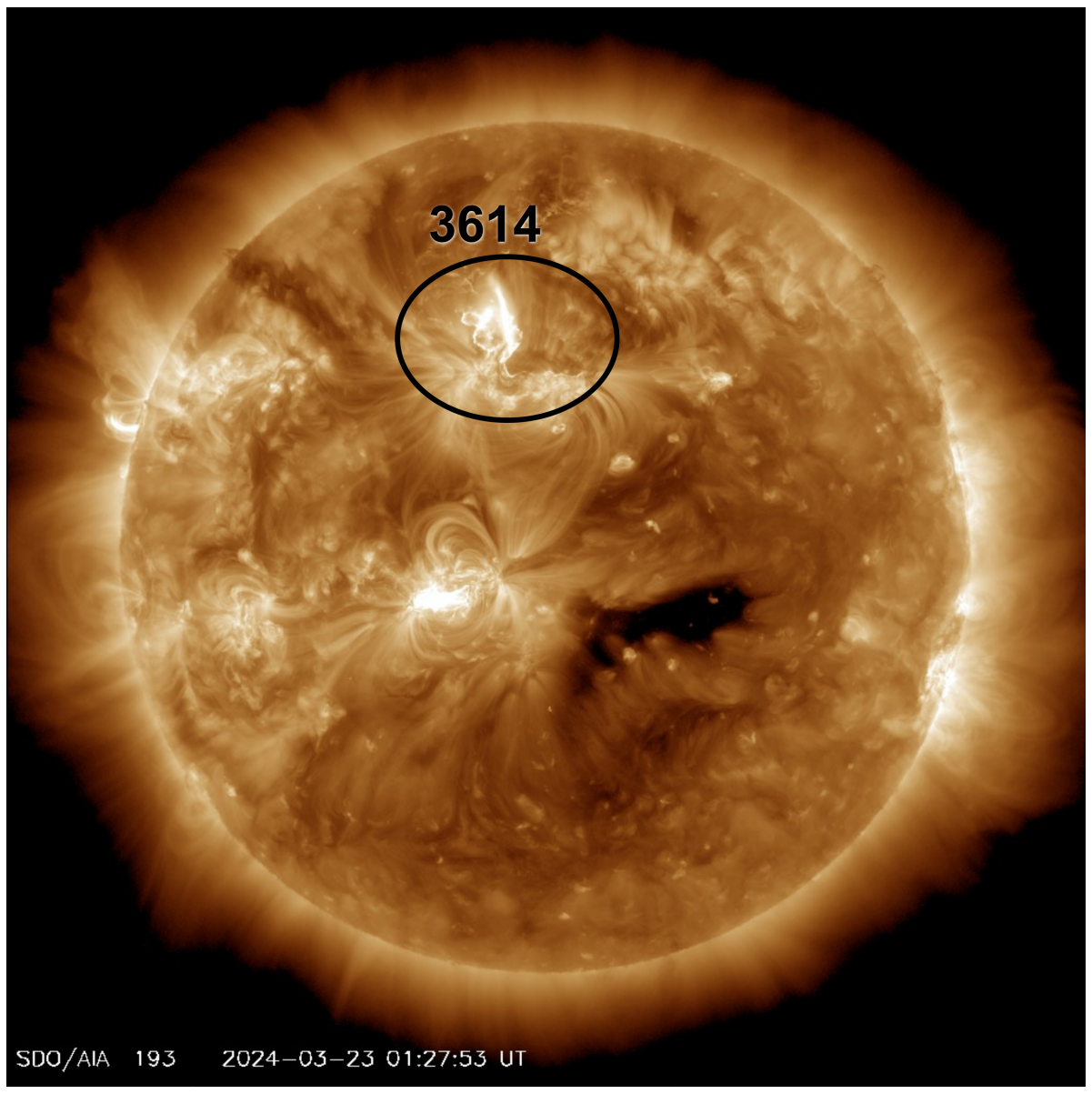

Figure 12: Active region 3614 originating the flare and the CME accelerating particles observed with Solar Orbiter on March 23, 2024*.

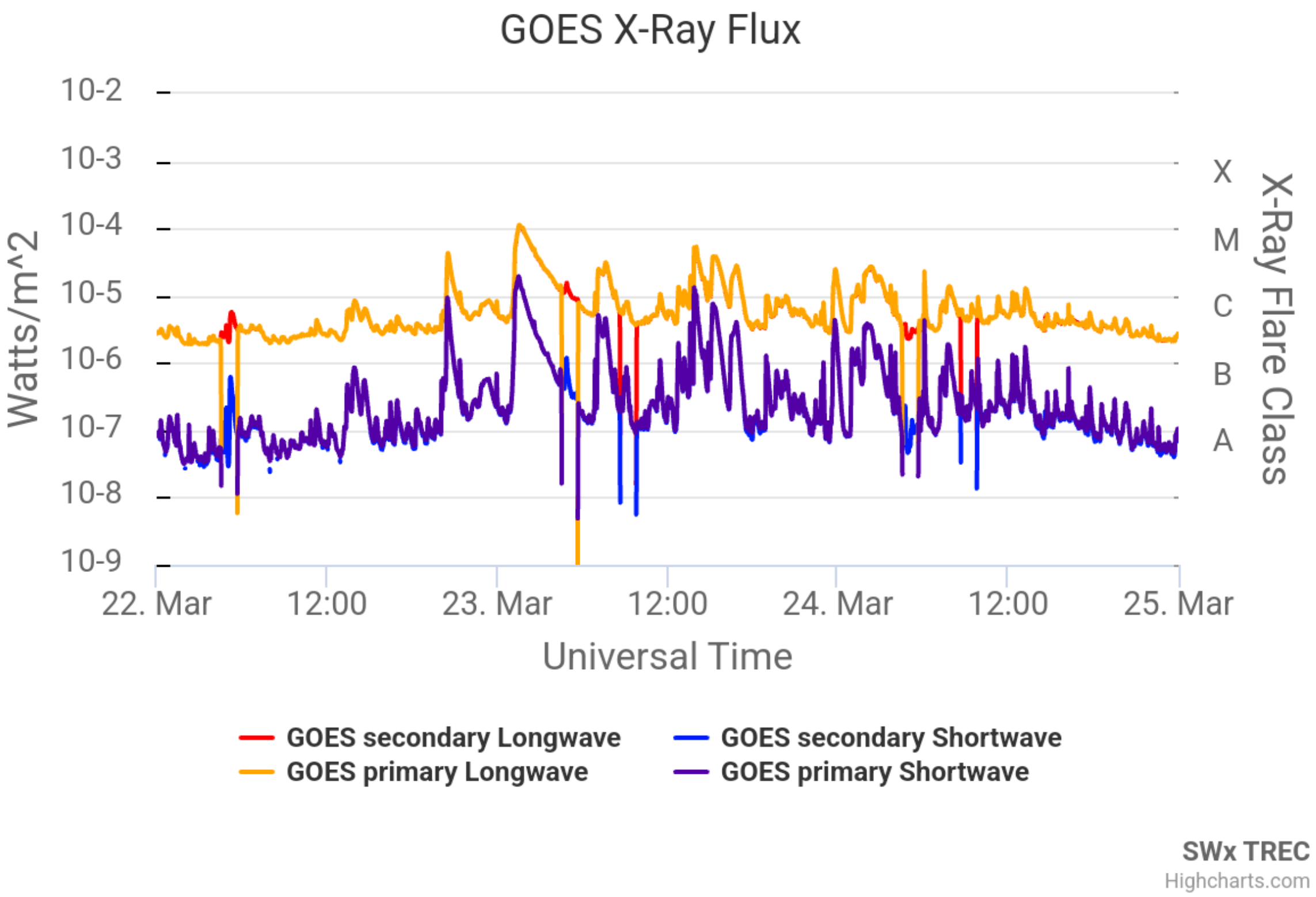

Figure 13: X-ray emission associated with the flare X1.12, observed by GOES-18.

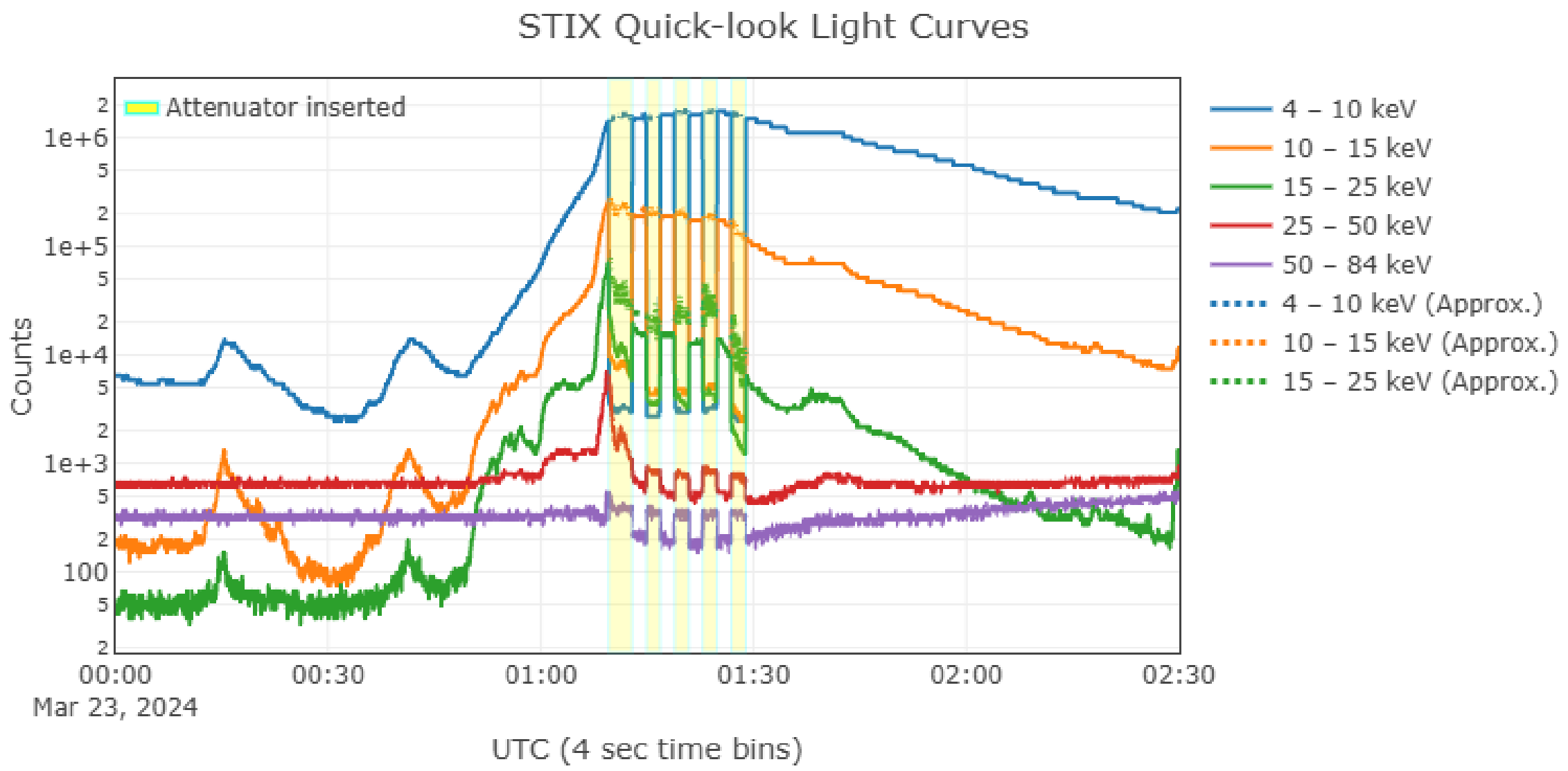

Figure 14: Same as Figure 13 for the STIX X-ray data observed in the early morning of March 23, 2024. Data are binned to 4 seconds.

was emitted after the flare at 01:25:51 UT (Figure 15)††, plausibly from the same active region. A type II radio burst was detected after 01:08 UT by the ALASKA-COHOE and Australia-ASSA stations of the e-CALLISTO system, revealing that acceleration of particles had occurred at the shock of the CME, as shown in Figure 16.

Electrons arrived at the Solar Orbiter spacecraft slightly before protons due to velocity dispersion. As shown in Figure 17, a sudden increase in the electron flux was observed at 01:30 UT in the energy range 50-100 keV, about 45 minutes before that of protons in the energy range 80-90 MeV and approximately half an hour after the X-ray enhancement.

### *4.2. Solar energetic protons*

As mentioned above, the increase of the overall proton flux above the GCR background, associated with the SEP event considered here as a case study, was observed with the EPD/HET instrument aboard Solar Orbiter after 02:00 UT. This event was characterized by the largest proton flux increase above 80 MeV from the launch of the mission in 2020 until March 2024. We are primarily interested in this energy range characterized by protons and nuclei that contribute to the deep charging of large-mission satellites and to dose absorption on manned missions.

The event onset, defined as the time when the proton flux exceeded the GCR

††`https://cdaw.gsfc.nasa.gov/CME_list/index.html`
movie webpage
`https://soleil.i4ds.ch/solarradio/data/BurstLists/2010-yyyy_Monstein/`,
`http://soleil.i4ds.ch/solarradio/callistoQuicklooks/?date=20230224`

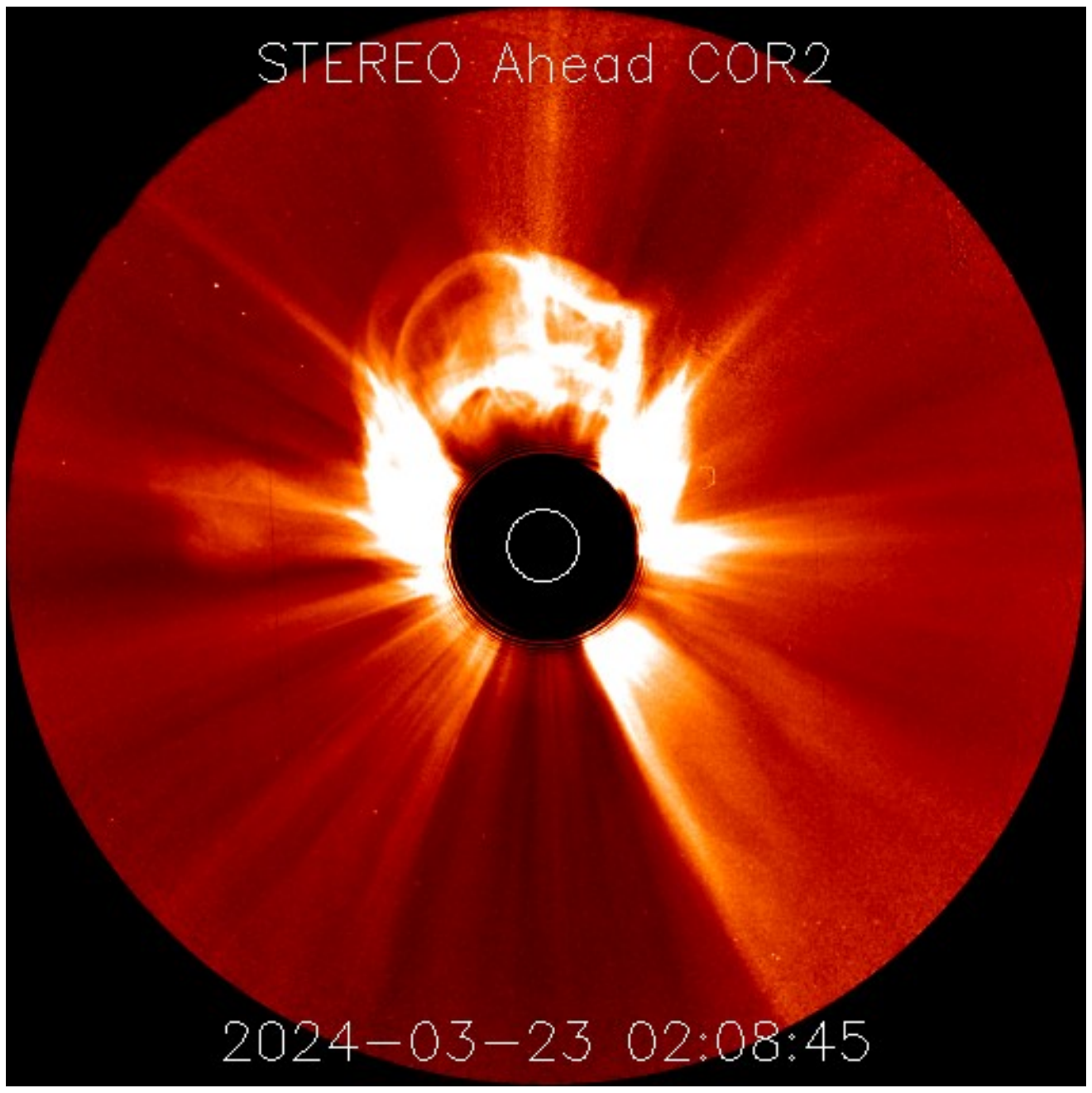

Figure 15: Coronal mass ejection as it entered the field of view of Stereo A COR2.

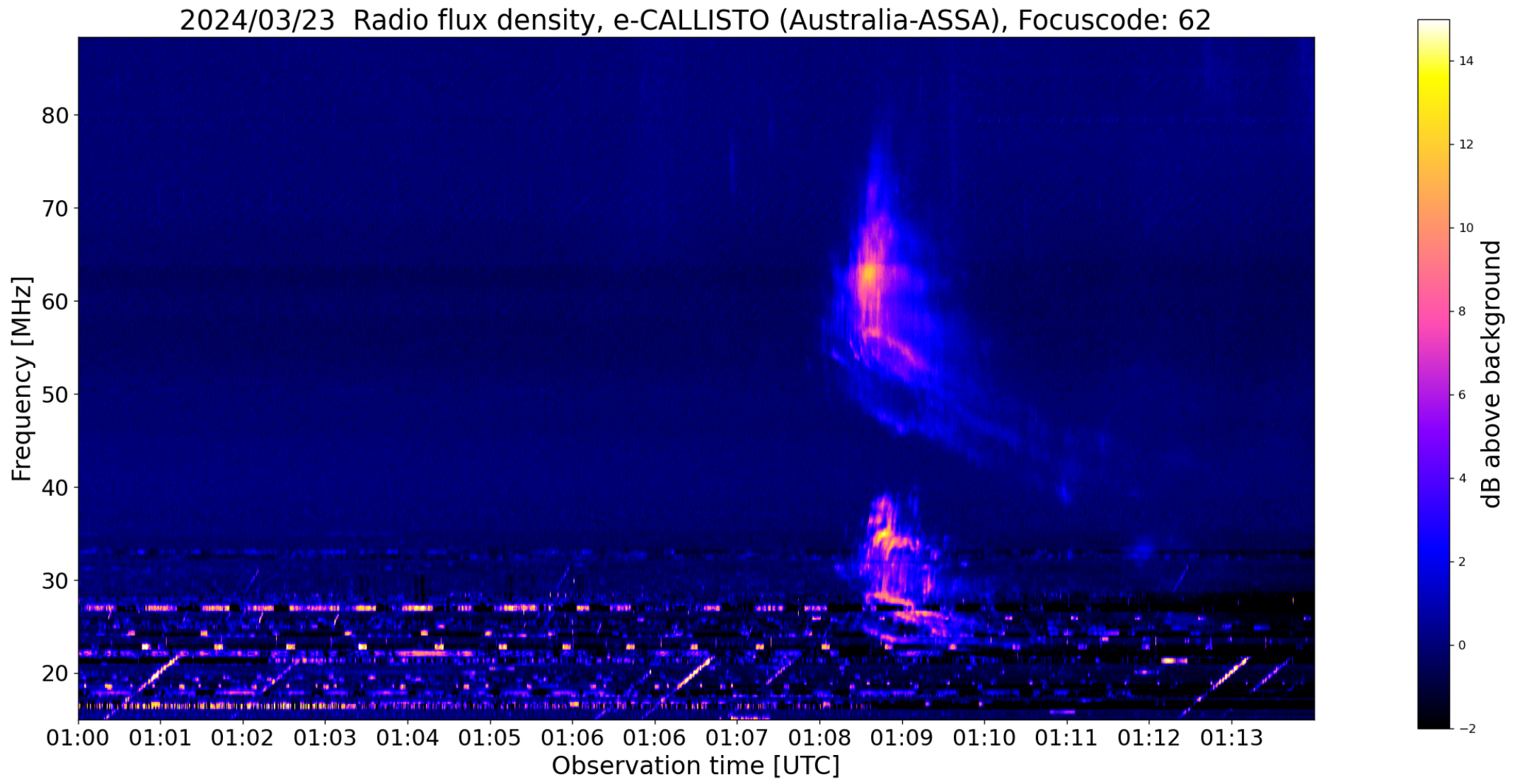

Figure 16: Type II radio burst emission detected by the e-Callisto system at the Australia-ASSA station attributable to the SEP event started on March 23, 2024.

background by 5 $\sigma$, was set between 02:45 and 03:15 UT above 80 MeV on board Solar Orbiter. The proton flux at the onset and peak of the SEP event was compared in Figure 8 to the homologous phases of other events of the same fluence intensity class ($10^7$–$10^8$ protons $cm^{-2}$). It is observed that this SEP event has a quite modest onset flux and a large peak flux, suggesting that Solar Orbiter was not magnetically well connected to the active region of the Sun at the origin of the particle acceleration.

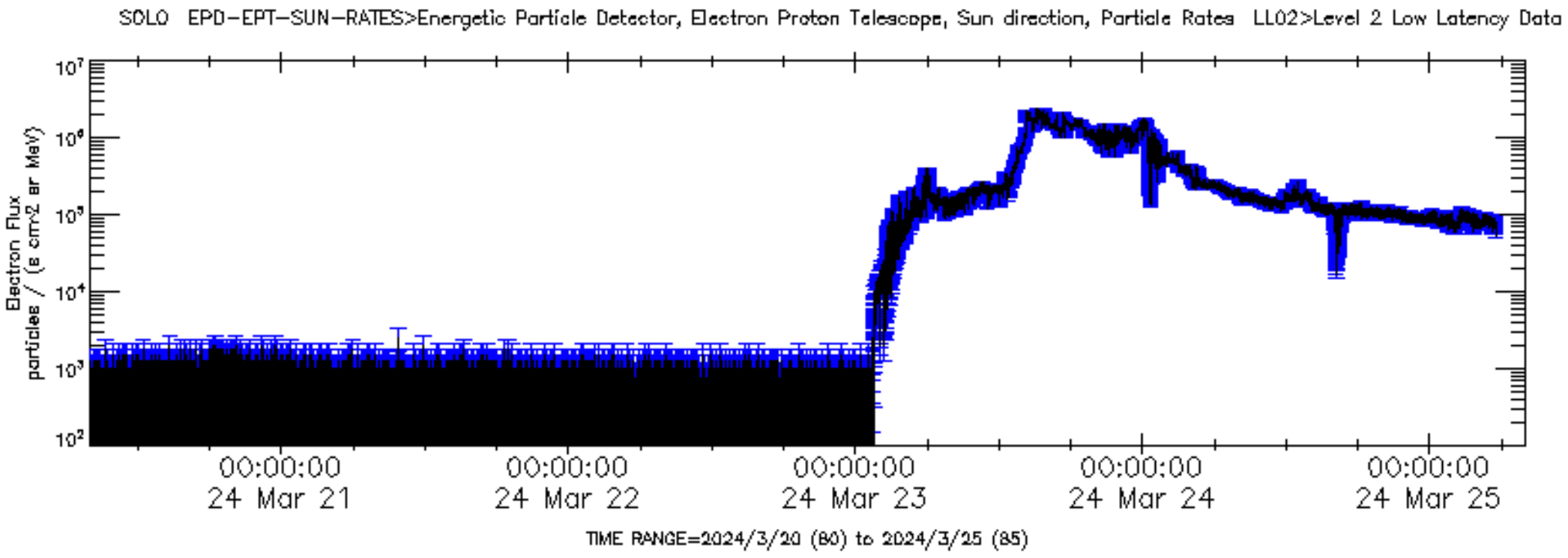

Figure 17: Electron flux enhancement between 50 and 100 keV observed with EPD/EPT on board Solar Orbiter on March 23-25, 2024. The onset is observed at approximately 01:30 UT.

## 5. Expected HASPIDE-SPACE performance during the March 23, 2024 event

The performance of HASPIDE-SPACE is estimated in the following with the FLUKA Monte Carlo. In order to follow the entire dynamics of solar activity leading to the SEP acceleration, we are interested in the detection of photons, electrons and protons. In Grimani et al. [22] we focused on the performance of individual a-Si:H sensors developed for medical applications by considering the average and standard deviation of the particle energy deposit distributions in their sensitive part. This can be considered the worst-case scenario for particle detection. As a matter of fact, due to the small thickness of our sensors, the deposits in the tail of the energy-loss distributions play a crucial role in determining the instrument actual capability to detect and monitor solar eruptions and SEP events. In Section 2 we showed that photon, electron, and SEP detection would be feasible with a signal-to-noise ratio of 5 if 60 MeV of energy per second is deposited in the new sensors, compared to 105 MeV in sensors for medical applications.

Comparison of previous and current Monte Carlo simulations allowed us to demonstrate that the sensor geometry optimized for proton detection is also feasible for X-ray and electron observations to monitor the entire evolution of solar activity leading to particle acceleration. Due to the saturation of the STIX instrument and the lack of availability of energy differential flux of photons, we are not able to simulate the X-ray photon flare observed early on March 23, 2024. However, based on the data appearing in Figure 14, $10^6$ counts in 4-second time bins constitutes a lower limit on the actual flux of photons that reached the instrument at the peak of the event. Based on table 4 of [22], which considered smaller and noisier sensors, a photon count rate > $10^4$–$10^5$ $s^{-1}$ in the 50-100 keV energy range would have sufficed to detect the flare. This indicates that HASPIDE-SPACE, in its current configuration, would have detected the flare with a higher signal-to-noise ratio. It is worth emphasizing that the detection of flares is allowed, as in this case, if the active region of the eruption lies in the field of view of the instrument.

In Figure 17, where the electron flux is reported in the energy range between 50 and 100 keV, an increase of more than four orders of magnitude is observed compared to the flare enhancement of the solar electron flux considered in [22]. To our knowledge, the differential energy flux of electrons during the evolution of the March 23, 2024 event is also not available. However, following the discussion reported in our previous paper for electron detection with smaller and noisier sensors and based on Figure 17, we would be able to monitor the flux of electrons for the entire duration of this event. Flare photons and ICME shock-accelerated electrons reach instruments in space before protons due to velocity dispersion. The detection of these particles can be disentangled from the arrival of protons since solar photons and electrons would only penetrate the first passive plane of HASPIDE-SPACE, while protons could also pass through the other passive layers of the instrument. Photon and electron detections are mainly meant for event dynamics monitoring, space weather investigations and solar proton enhancement nowcasting, since these particles are the main contributors to satellite internal charging and dose absorption.

Our main goal remains the detection of protons up to hundreds of MeV, although we also monitor protons at lower energies. The SEP case study considered here was associated with the observation of the acceleration of protons up to 1 GeV. At this energy the solar proton flux appears several orders of magnitude above the GCR background at the peak of the event. Therefore, this high-energy event would have generated signals deeply into the instrument. The results we are presenting are based on Monte Carlo simulations of proton energy losses per second during the different phases of the March 23, 2024 event, made possible by the availability of absolute differential proton fluxes measured on board Solar Orbiter. At the peak of the event, these results constitute a lower limit since the proton flux measurements were not available above 1 GeV. In the simulations, we examined different particle flux incidence directions on the instrument, starting with perpendicular incidence on plane A (front incidence; Figure 18 (a)) at the onset, assuming the instrument being magnetically well connected to the Sun active region responsible for proton acceleration.

In Figure 19 it can be observed that at the onset of the event a large amount of energy is deposited in the first active layer A and in the top side sensors (F1, F2, F3, F4), thus indicating that during this phase, protons are either stopped in the first passive layer of tungsten or escape the system, while the high-energy part of the flux is not intense enough to activate the bottom sensitive planes. The real direction of incident particles would be revealed by an appropriate combination of particle energy losses in the side sensors and energy losses in the sensitive planes. At the peak of the proton event between 14:15 UT and 15:15 UT photons are totally depressed, the electron flux is at the peak but only the first plane of the detector is activated by these particles. Protons penetrate the instrument down to the sensitive layer C with energy deposits well above the sensitivity of the instrument, with side sensors F1-F4 indicating that

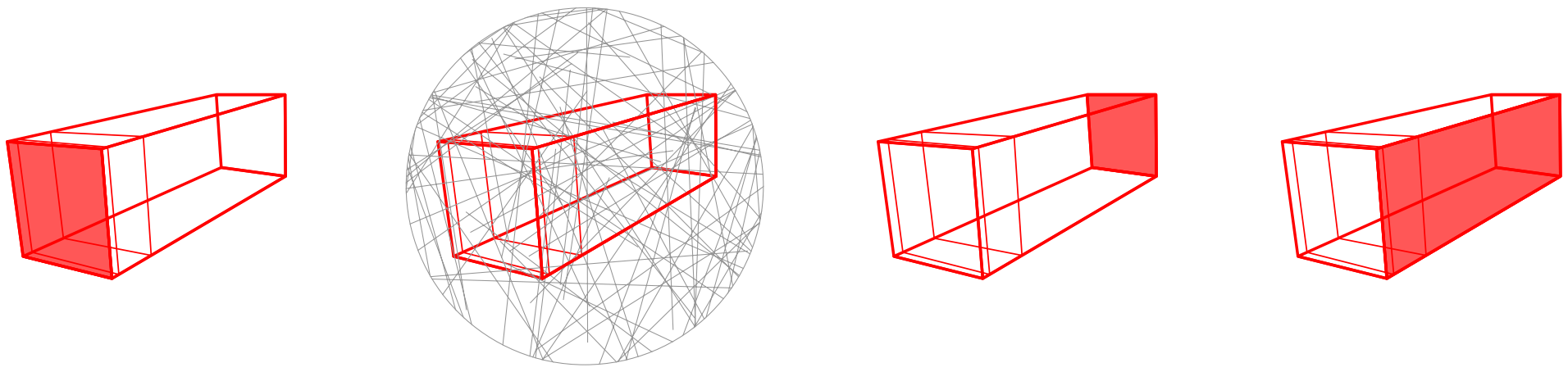

**(a)** **(b)** **(c)** **(d)**

Figure 18: Sketch of particle flux incident on the detector with different incidence directions. From left to right: (a) particle perpendicular incidence on layer A (front incidence); (b) particle isotropic incidence on the instrument; (c) particle perpendicular incidence on layer D (back incidence); (d) particle perpendicular incidence on the instrument side.

most of the protons are stopped in the detector in the thickest block of tungsten (see Figure 20). It is interesting to note that the proton flux is not intense enough above 500 MeV for the energy deposits of the particles to exceed the sensitivity of the instrument in layer D. A proton flux two orders of magnitude larger would be needed for detection. This observation reveals that our instrument is optimum for SEP detection above 500 MeV characterized by fluences of the order of $10^9$ protons $cm^{-2}$ or more above tens of MeV. Weaker events will be monitored at lower energies.

In Figure 21, we considered the same proton incidence on the A layer, but during the decay phase of the event. During this phase, the distributions of proton energy losses are higher in the top layer A and in the side sensors F1-F4 with respect to the onset phase. It is worthwhile to recall that during the decay phase of SEP events the spatial distribution of protons tend to become isotropic (see Figure 18 (b); [5]). Therefore, in Figure 22 the decay phase of the event is simulated by considering an isotropic incidence of protons. The energy loss distributions, associated with particle isotropic incidence, overcome the instrument sensitivity in the layers A and D and in the side sensors.

Figure 23 illustrates the evolution of the March 23, 2024 event as plausibly observed by our prototype instrument HASPIDE-SPACE if it had been placed on board Solar Orbiter. Photons would have activated the first detecting plane between 00:50 UT and 02:00 UT, contributing also to the overall energy deposition in this plane during successive time intervals, as shown in the first panel of Figure 23. Electrons are expected to activate the same plane at 01:30 UT for approximately two days, as shown in the second panel of Figure 23. Proton flux enhancement would be observed at 02:45 UT (third panel of Figure 23). In the fourth panel, the time evolution of all particle detection capability with HASPIDE-SPACE is summarized.

For completeness, we also performed a full set of simulations for all phases of the March 23, 2024 event, assuming an isotropic particle distribution, perpendicular incidence on the layer D of the instrument (Figure 18 (c)) and perpendicular incidence on the side of the instrument (Figure 18 (d)). The simulation results are provided in

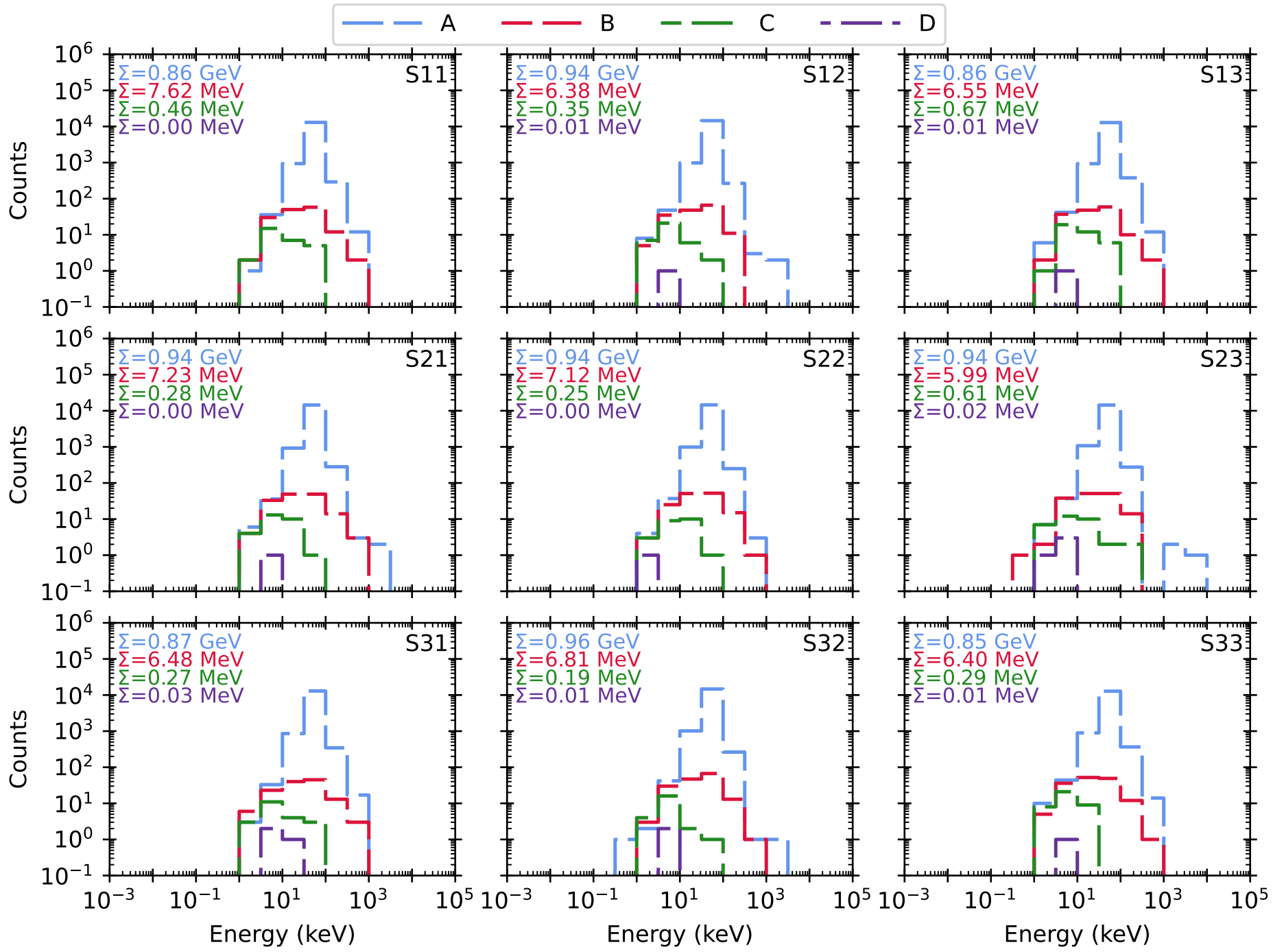

(a) Matrix sensors.

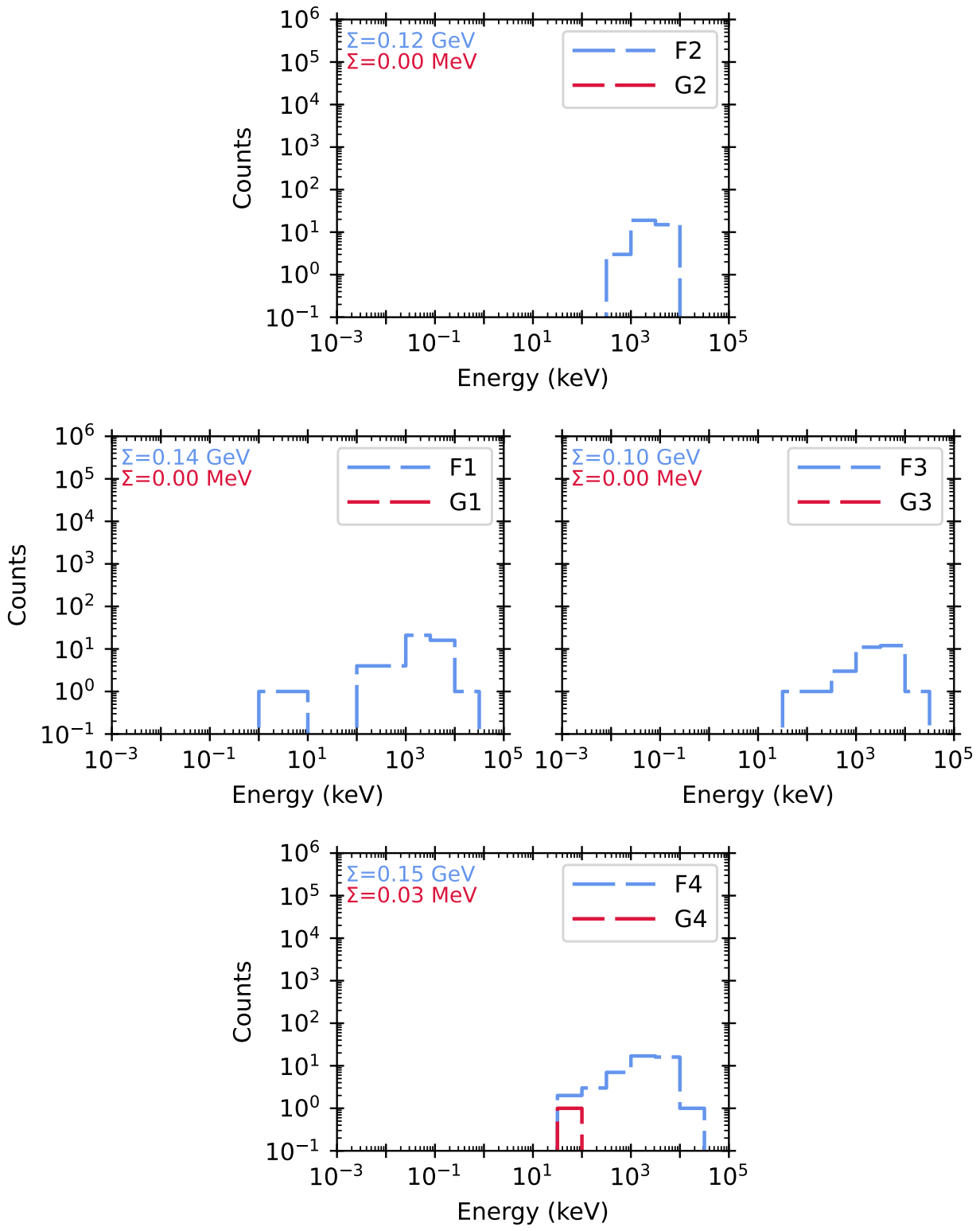

(b) Side sensors.

Figure 19: Total ionization energy loss deposited in 1 s in each sensor of each plane of the HASPIDE-SPACE instrument, as indicated in Figure 5. The position of the sensitive layers were also indicated in Figure 5. The plots correspond to the simulation of the March 23, 2024 SEP event at the onset, with particle perpendicular incidence on layer A as depicted in Figure 18 (a).

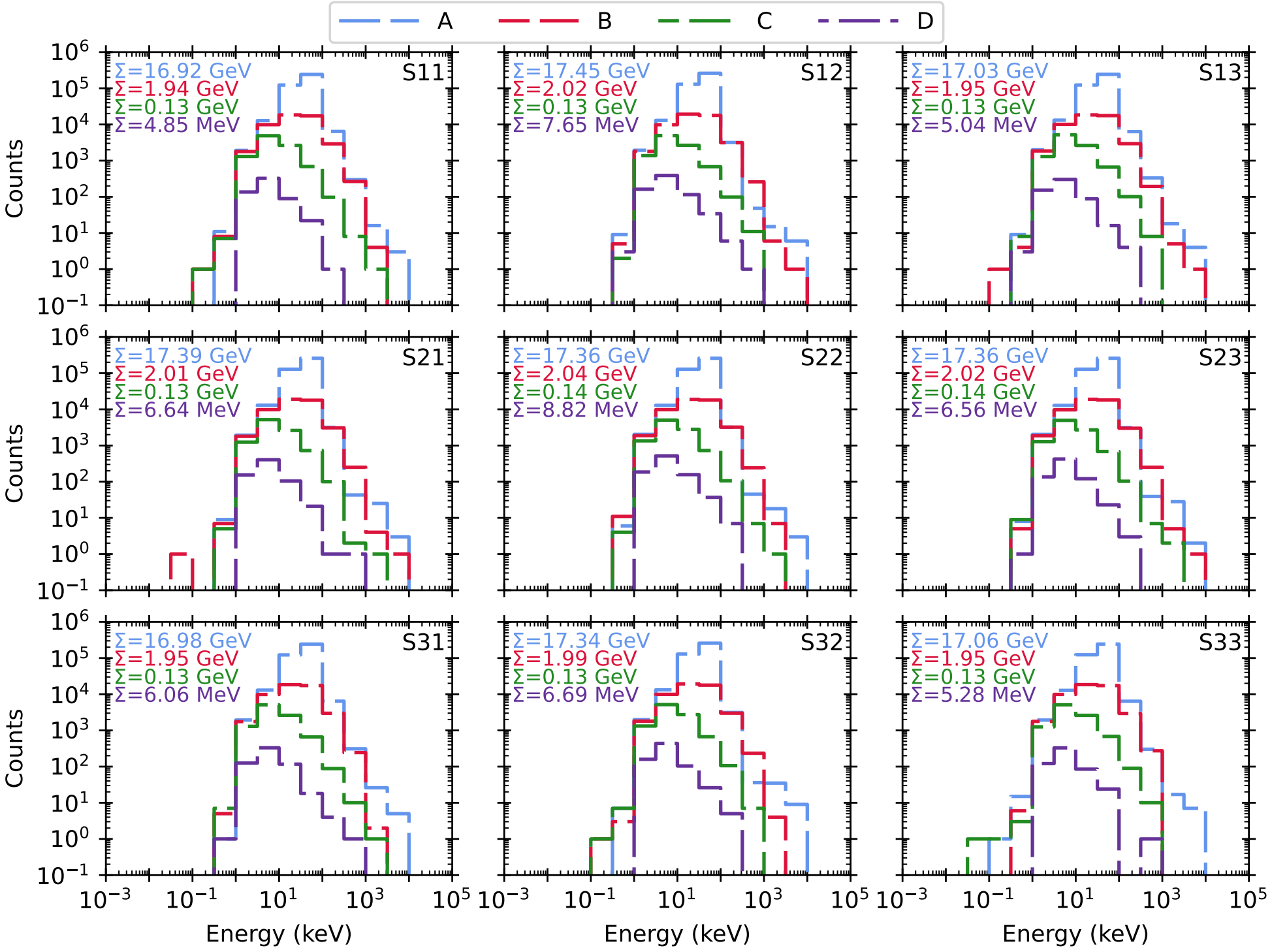

(a) Matrix sensors.

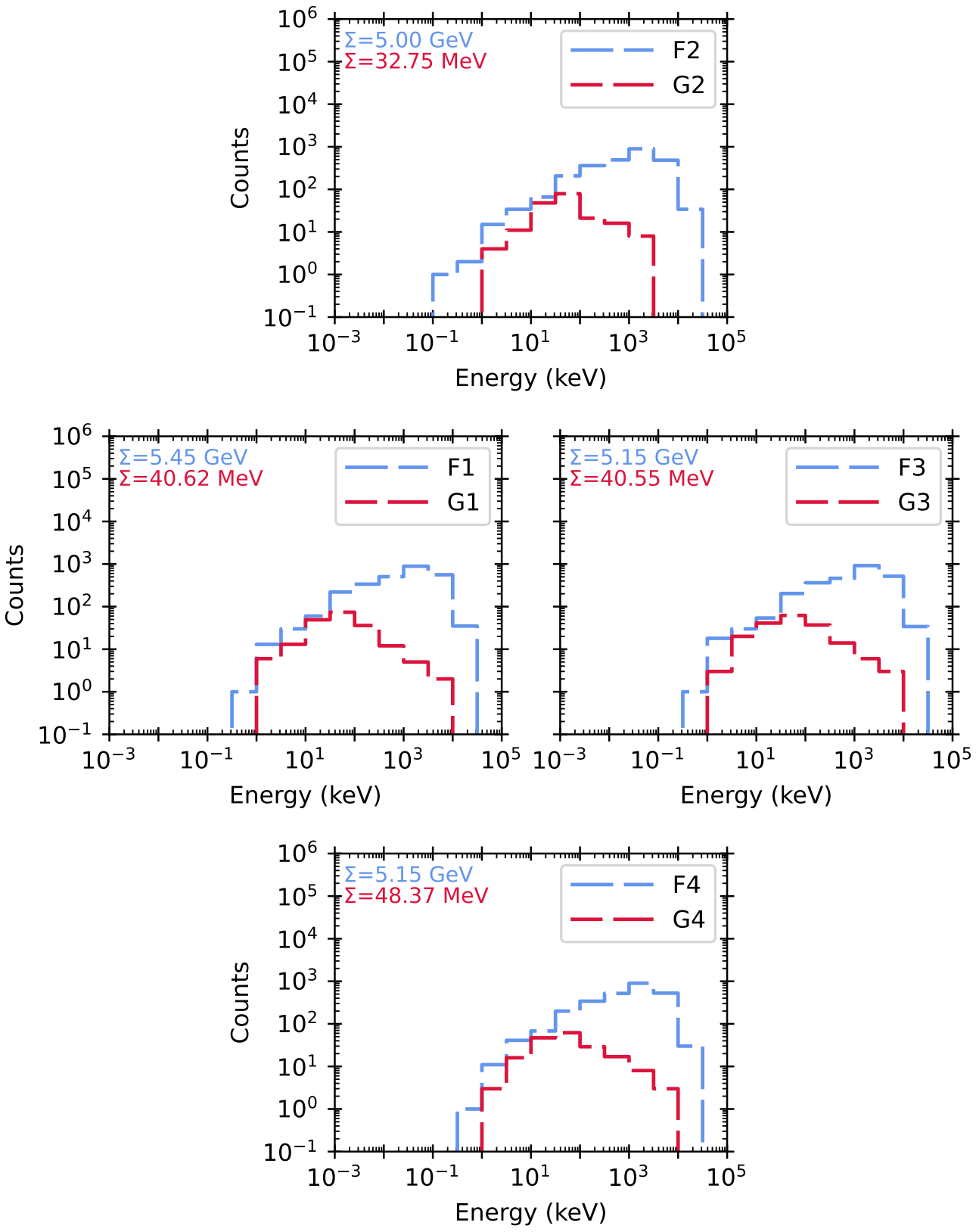

(b) Side sensors.

Figure 20: Same as Figure 19 for the event peak.

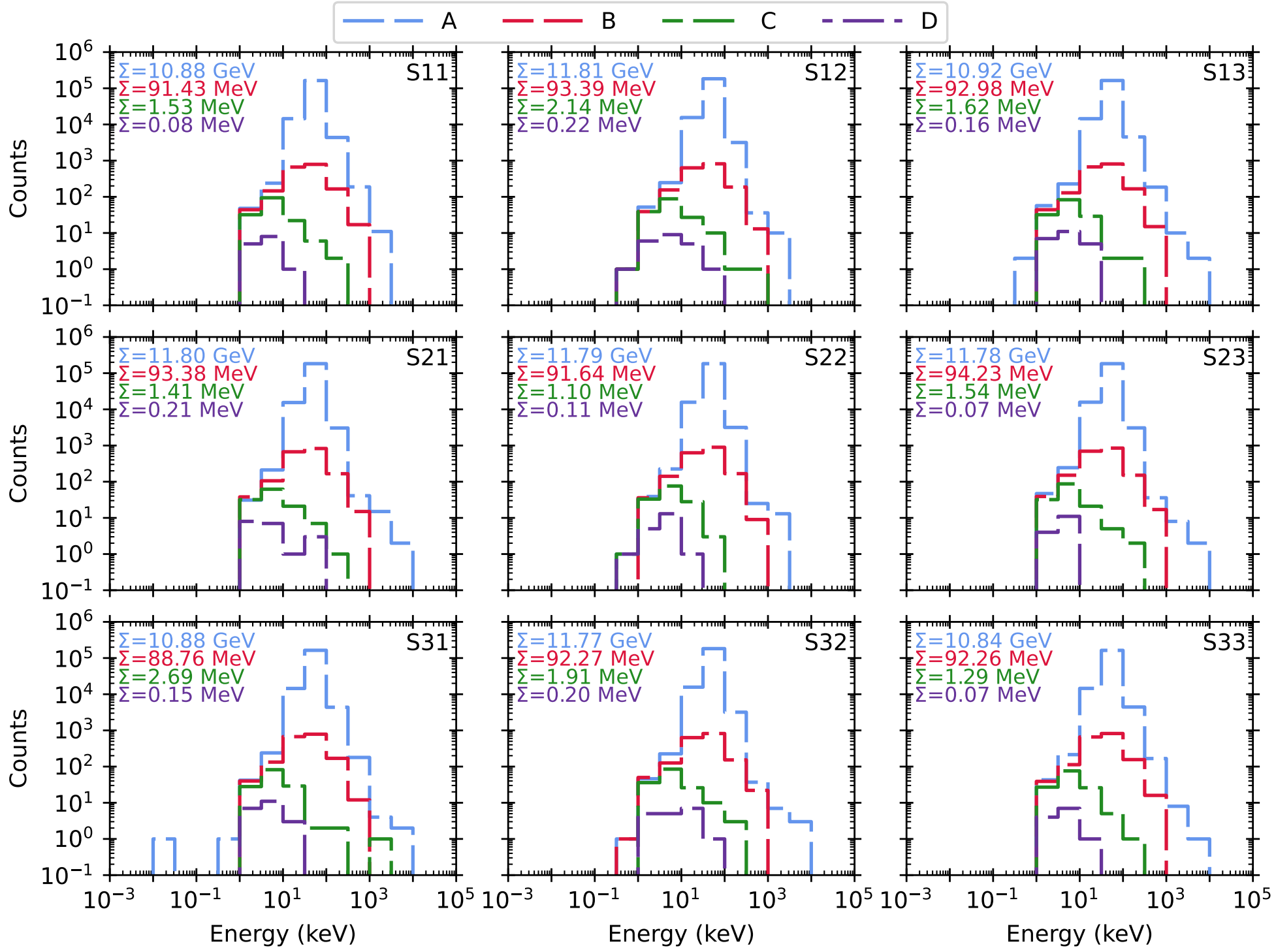

(a) Matrix sensors.

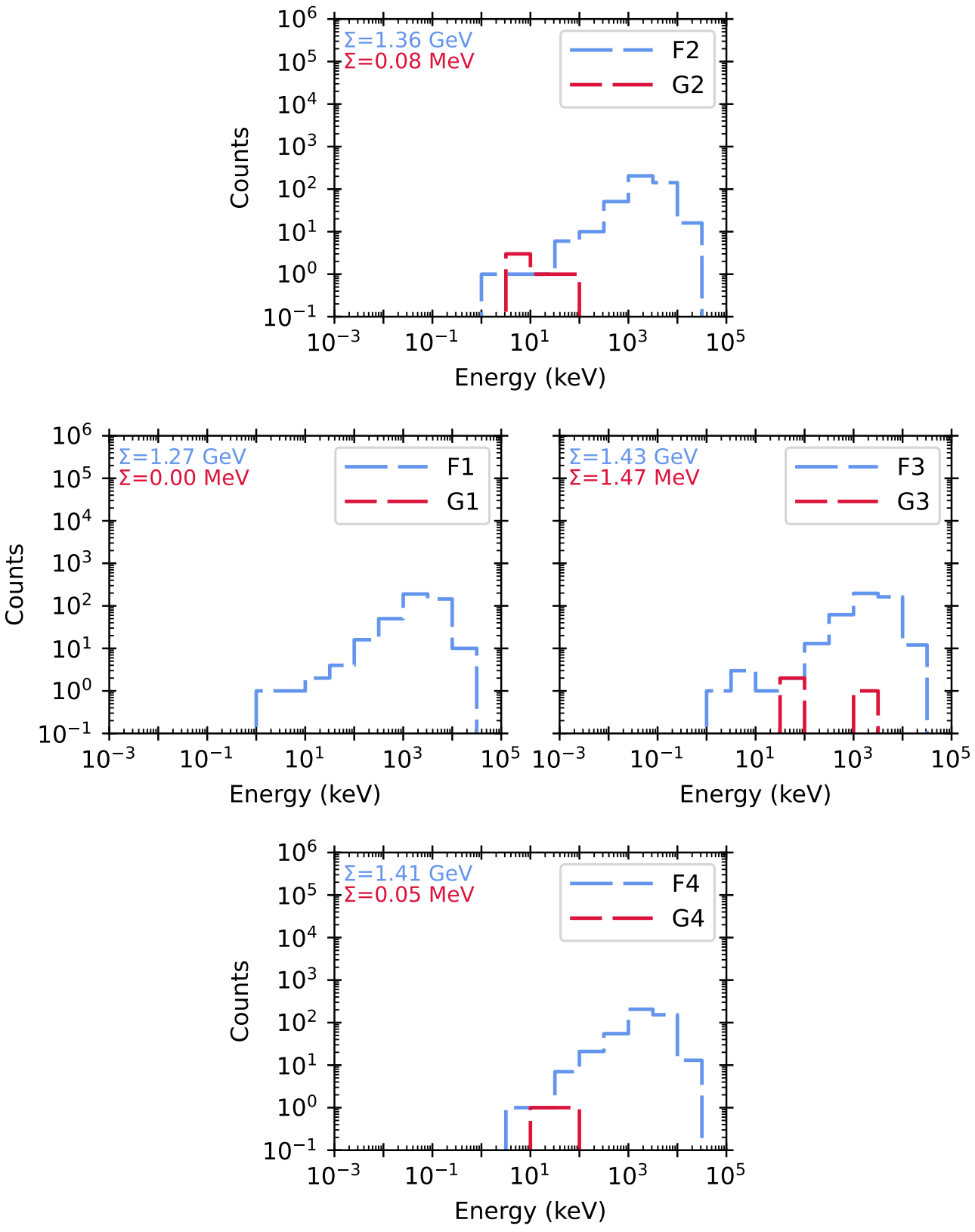

(b) Side sensors.

Figure 21: Same as Figure 19 for the event decay.

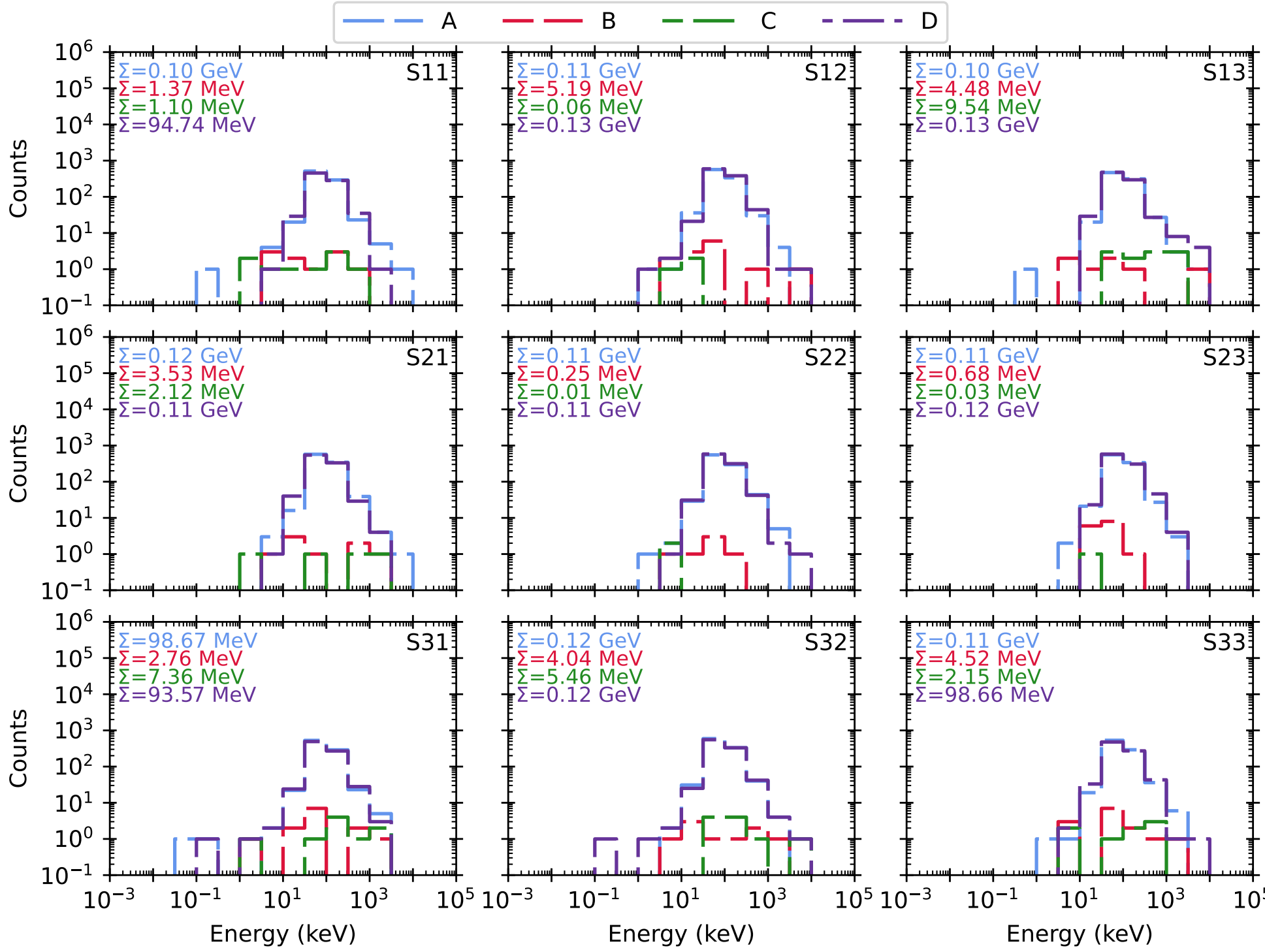

(a) Matrix sensors.

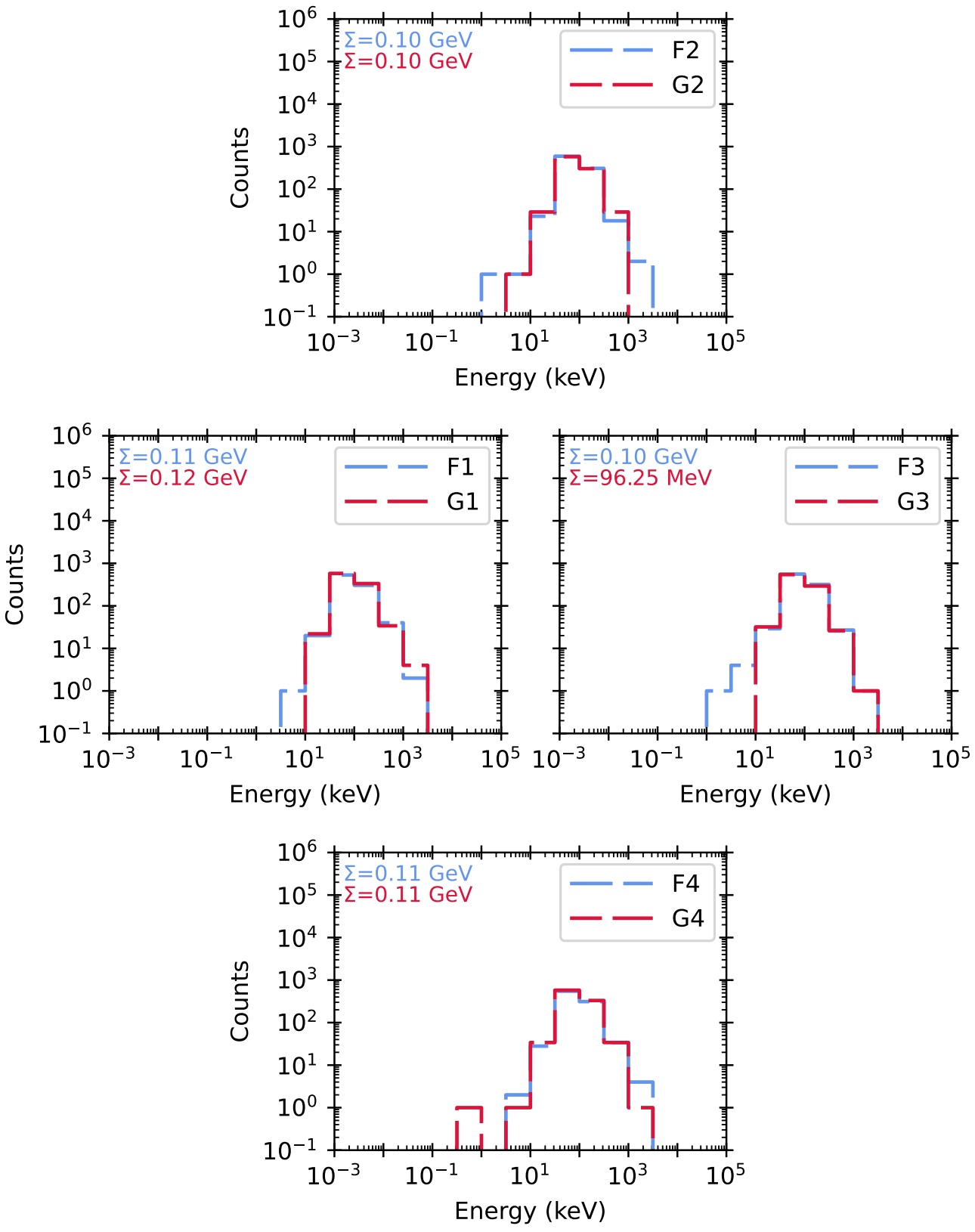

(b) Side sensors.

Figure 22: Same as Figure 21 with isotropic incidence (see Figure 18 (b)).

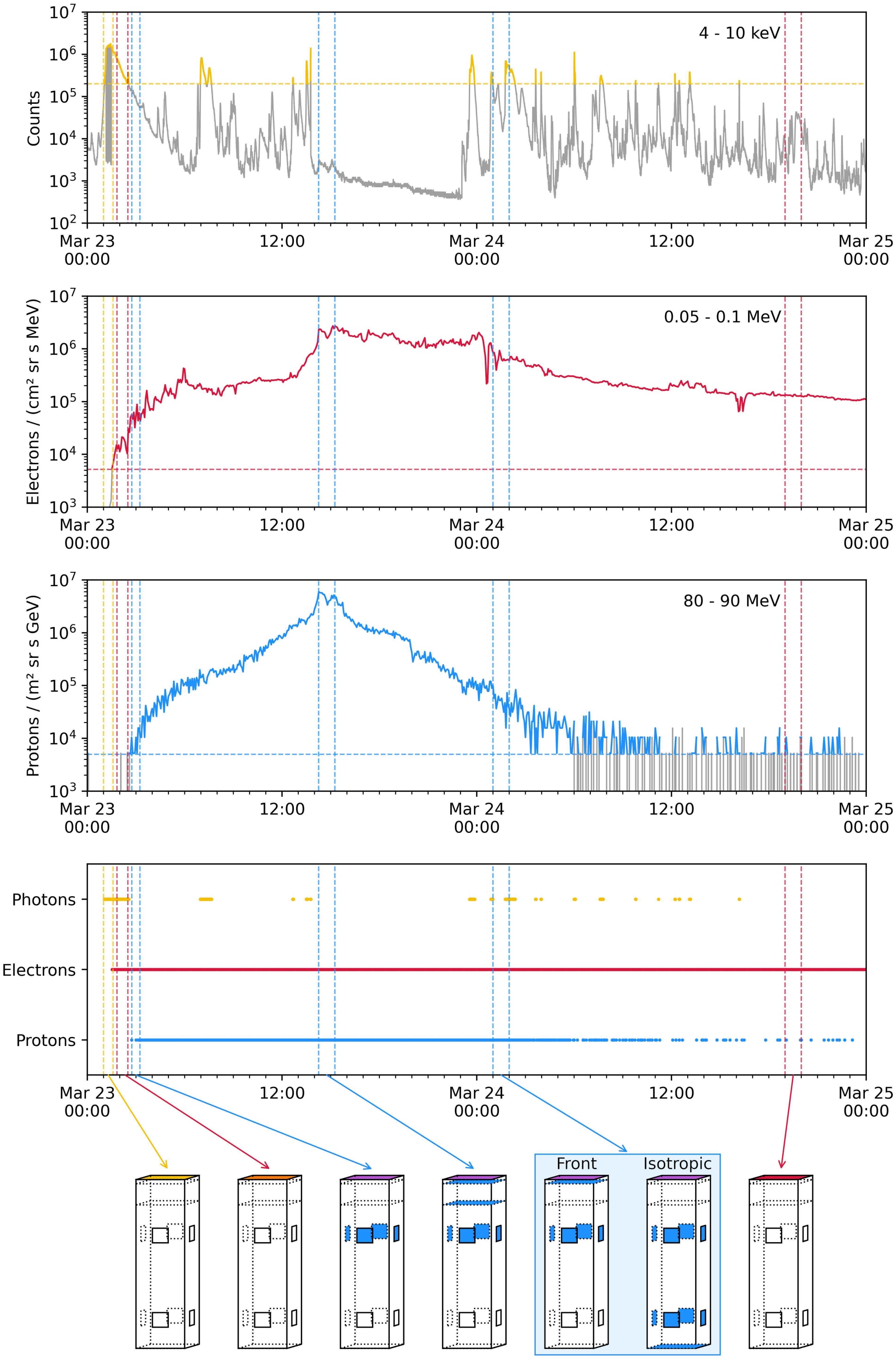

Figure 23: Solar eruption and SEP monitoring observed with HASPIDE-SPACE during the evolution of the March 23, 2024 event with particles incident on the top layer (plane A) of the instrument. For the decay phase of the event, both front and isotropic incidence of the particles were considered.

the supplementary material accompanying this paper.

Finally, in Figures 24 to 26 we report the energy deposited in one second on the a-Si:H sensors when the angle of incidence of the protons arriving from one side of the instrument varies in steps of 15 degrees from 0 (corresponding to the normal incidence on one side of the tower, as in Figure 18 (d)) to 90 degrees (particle perpendicular incidence on layer A, as in Figure 18 (a)). These simulations will constitute a fundamental aid to support the data analysis activity when the instrument will be in space and SEPs will modify their spatial distribution during the evolution of each event.

## 6. Conclusions

Galactic cosmic rays and SEPs are at the origin of spacecraft deep charging and astronaut absorbed doses. This work presents the design and performance of HASPIDE-SPACE, a radiation-hard prototype instrument developed for monitoring medium-intense solar activity and SEP events. HASPIDE-SPACE has been developed within the framework of the HASPIDE experiment funded by the Fifth Commission of the Italian Institute for Nuclear Physics. The instrument is tower-shaped and consists of four active layers made of 3x3 arrays of a-Si:H sensors with an area of $7\times7$ mm$^2$ and 10 $\mu$m thickness, specifically developed for this application. Three passive tungsten blocks are placed between the active planes for a total of 8 cm thickness. The instrument has a mass of 1.3 kg and overall dimensions of $2.4\times2.4\times12$ cm$^3$. We have discussed the ability of HASPIDE-SPACE to monitor the entire evolution of solar protons events with fluence larger than $10^7$ protons cm$^{-2}$ and the associated solar activity through the detection of photons and electrons flux monitoring. The instrument would also be capable of detecting the dynamics of protons above 500 MeV in case of SEP fluences above $10^9$ protons/cm$^2$, or up to 100-200 MeV in case of smaller fluence events. Our sensors would not be damaged or saturated by extreme fluence SEPs. Our instrument would be valuable for use aboard low Earth orbit missions. Finally, in the absence of solar activity, HASPIDE-SPACE could also play the role of strong long and short gamma-ray burst monitor.

## Acknowledgements

The authors are very grateful to Prof. Emanuele Lattanzi and Dott. Paolo Capellacci from the Department of Pure and Applied Sciences (DiSPeA) of the University of Urbino Carlo Bo for having allowed the use of the 3-D printer in their availability and for their support in printing the mechanical frames of the passive and active layers of the HASPIDE-SPACE prototype instrument.

## References

[1] H. M. Araújo et al. Detailed Calculation of Test-Mass Charging in the LISA Mission. *Astr. Phys.*, 22:451–469, 2005.

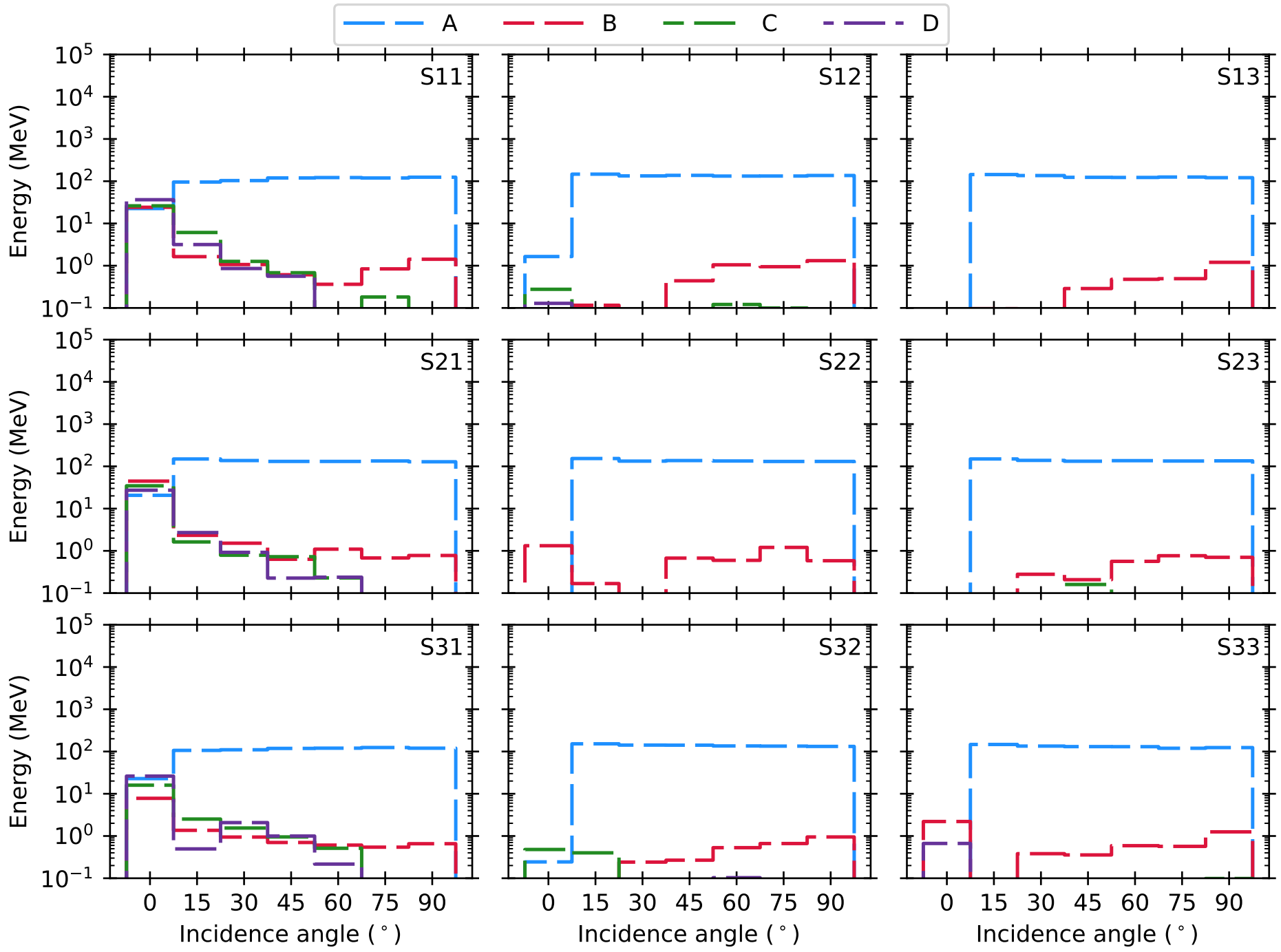

(a) Matrix sensors.

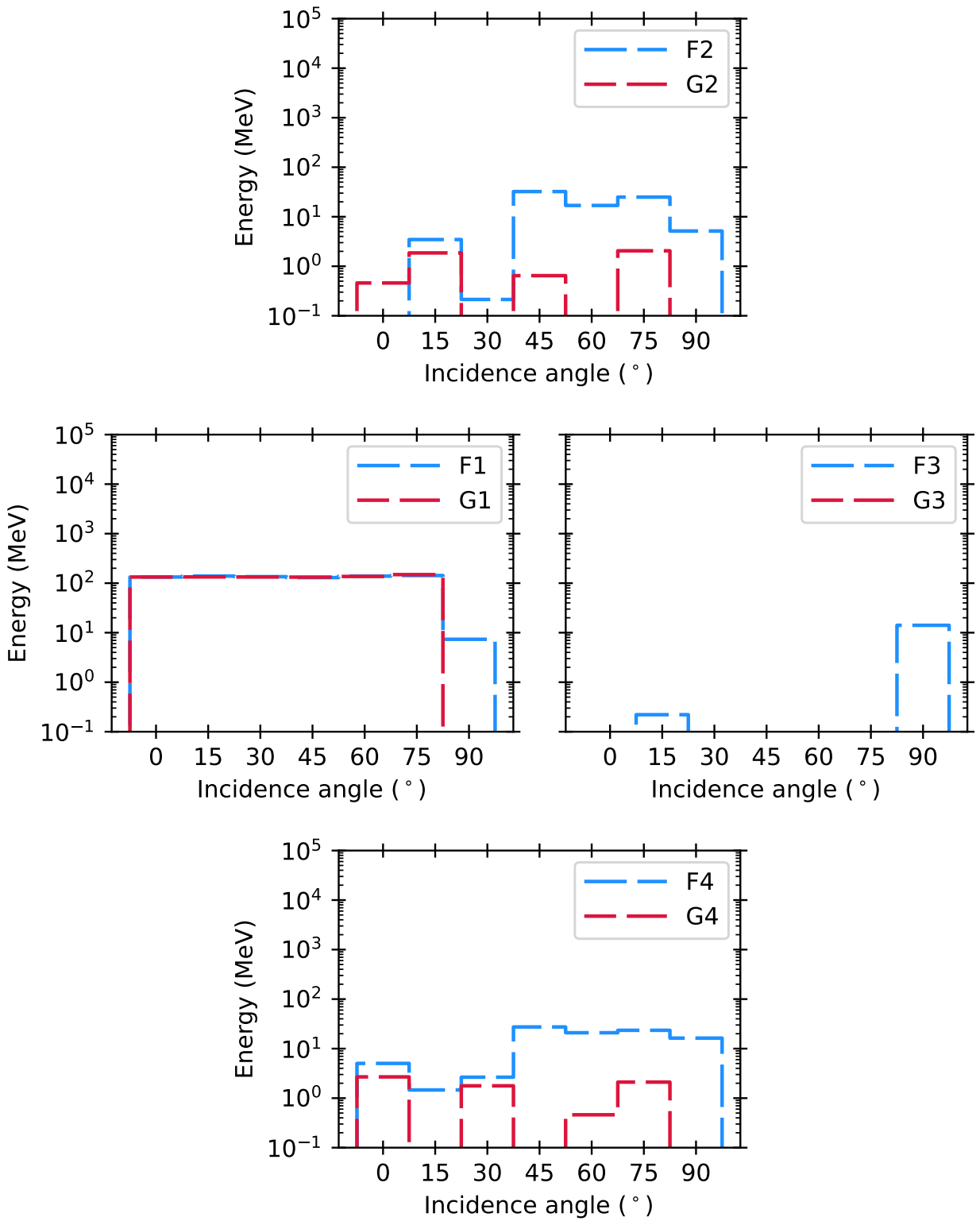

(b) Side sensors.

Figure 24: Onset of the March 23, 2024 SEP event with particle incidence on the instrument side from parallel through perpendicular incidence with 15 degrees step from the side area of the instrument, as indicated in Figure 18 (d).

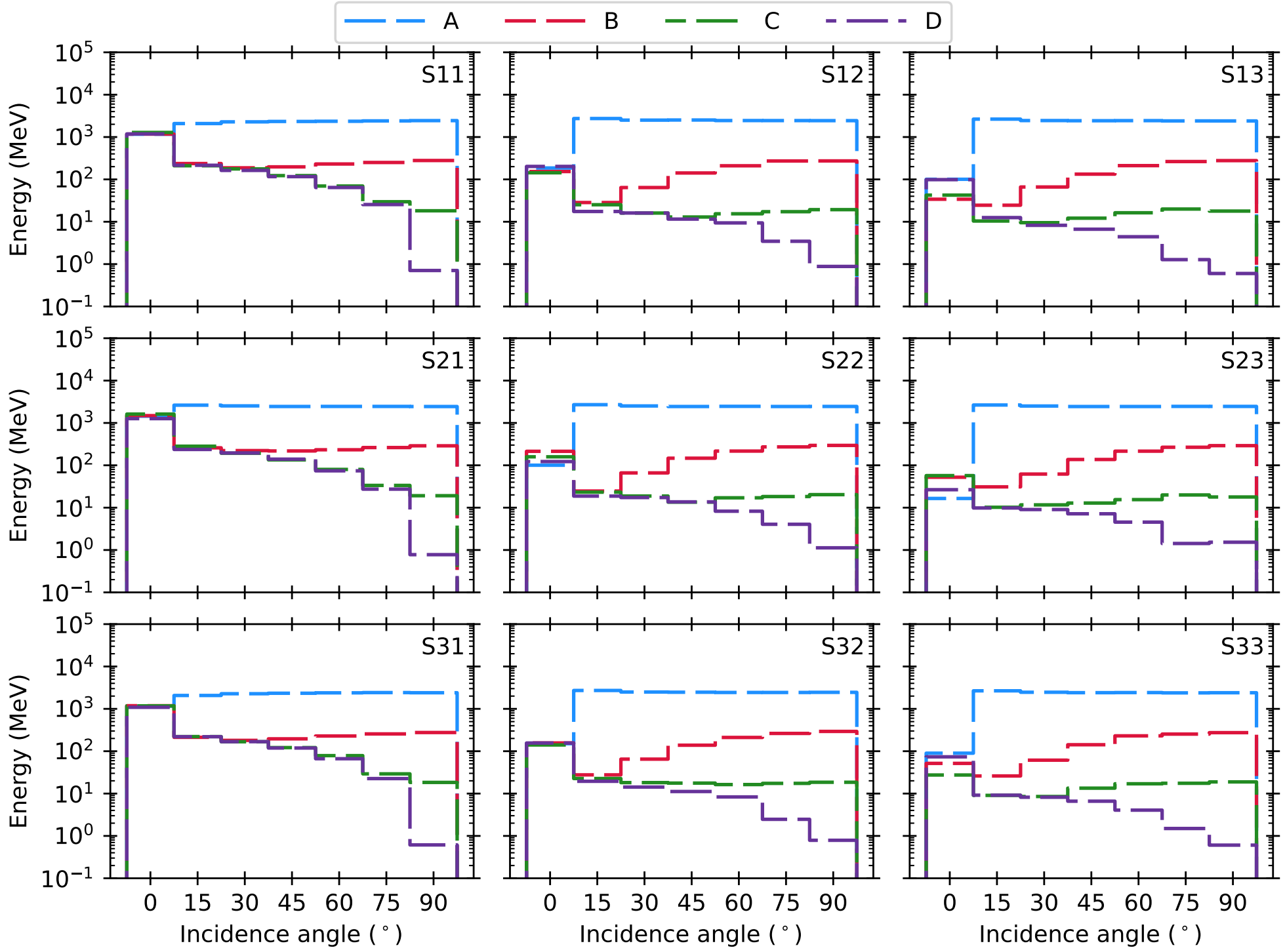

(a) Matrix sensors.

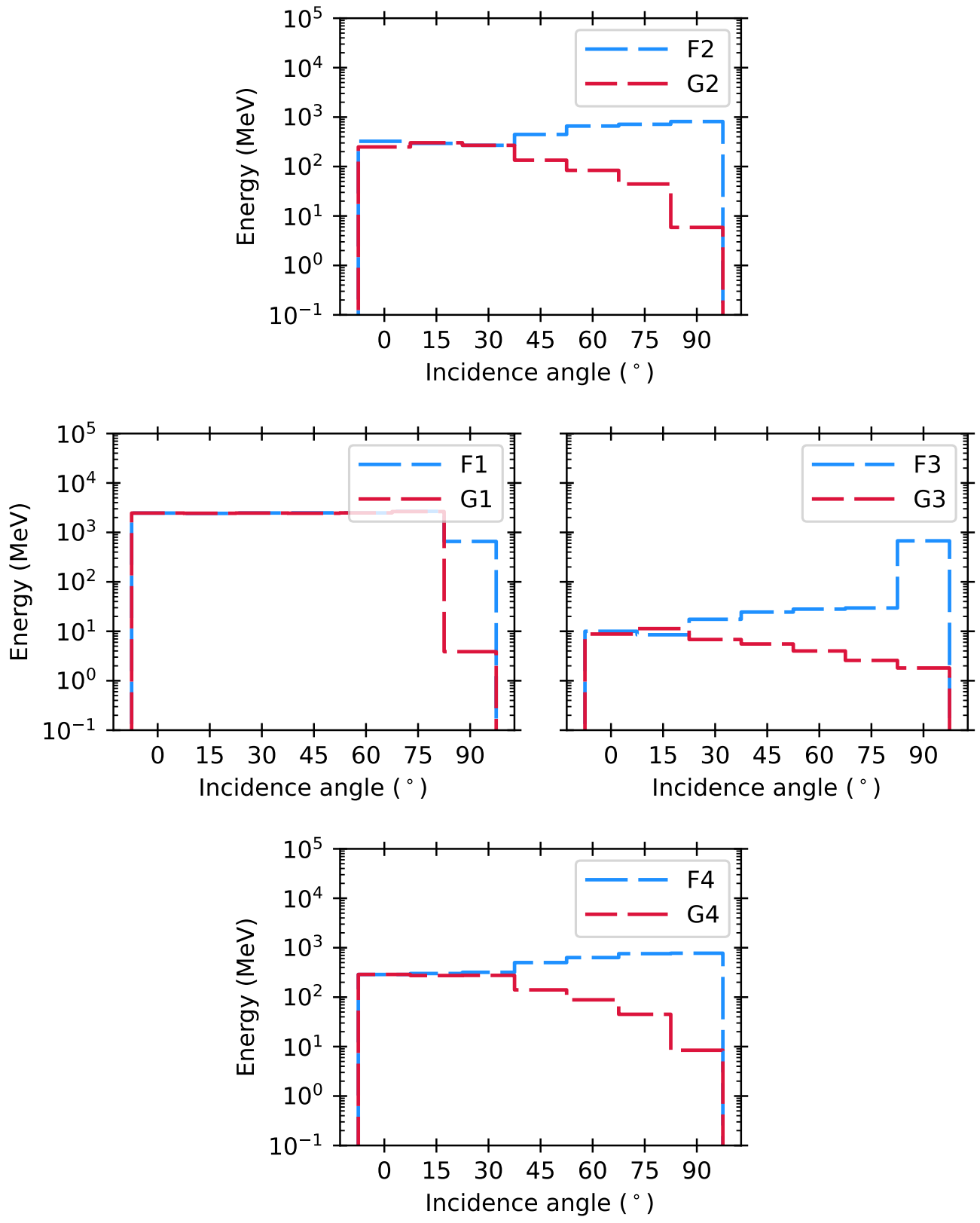

(b) Side sensors.

Figure 25: Same as Figure 24 for the SEP event peak.

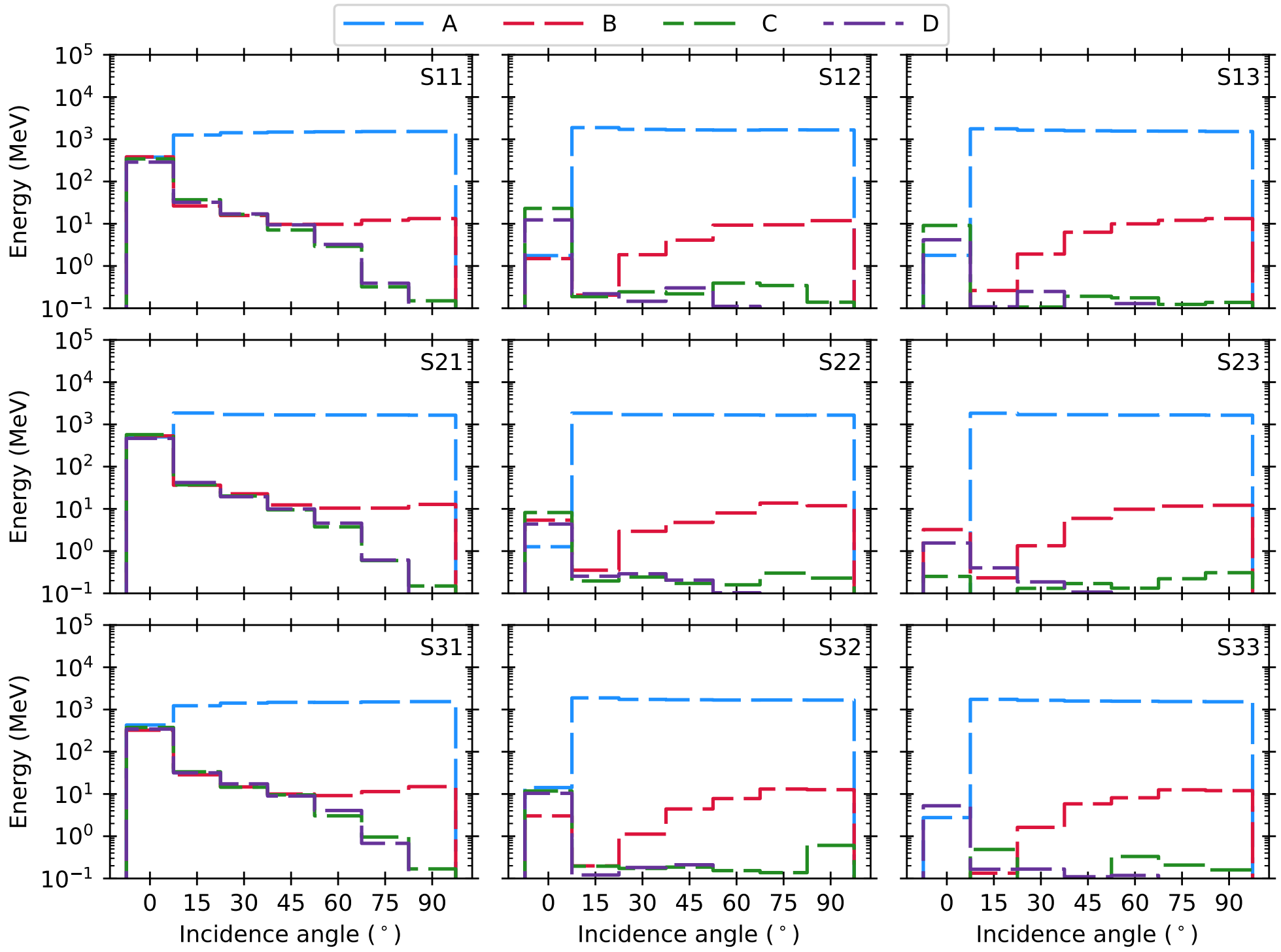

(a) Matrix sensors.

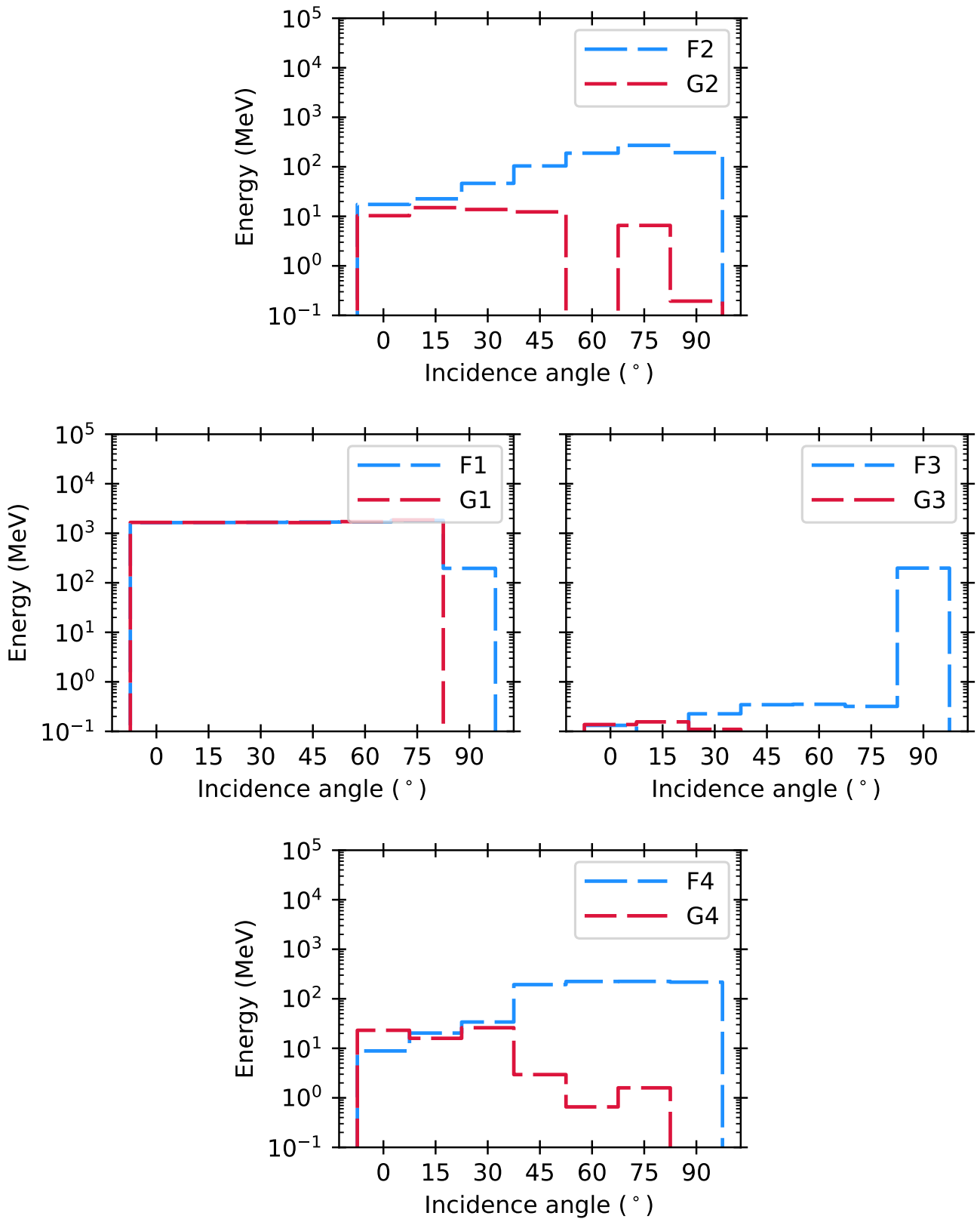

(b) Side sensors.

Figure 26: Same as Figure 24 for the SEP event decay.

[2] P. J. Wass, H. M. Araújo, D. N. A. Shaul, and T. J. Sumner. Test-mass charging simulations for the LISA Pathfinder mission. *Classical and Quantum Gravity*, 22:S311–S317, May 2005.

[3] Daniele Telloni, Michele Fabi, Catia Grimani, and Ester Antonucci. Metis aboard the solar orbiter space mission: Doses from galactic cosmic rays and solar energetic particles. *AIP Conference Proceedings*, 1720, 2016.

[4] C. Grimani, M. Villani, M. Fabi, A. Cesarini, and F. Sabbatini. Bridging the gap between Monte Carlo simulations and measurements of the LISA Pathfinder test-mass charging for LISA. *Astronomy and Astrophysics*, 666:A38, October 2022.

[5] C. Grimani, M. Fabi, A. Persici, F. Sabbatini, M. Villani, A. Burtovoi, E. Frassati, F. Pancrazzi, D. Telloni, P. Kühl, J. Rodríguez-Pacheco, R. F. Wimmer-Schweingruber, L. Abbo, V. Andretta, E. Antonucci, P. Chioetto, V. Da Deppo, Y. De Leo, S. Gissot, G. Jerse, F. Landini, M. Messerotti, G. Naletto, G. Nicolini, C. Plainaki, M. Romoli, G. Russano, C. Sasso, D. Spadaro, M. Stangalini, R. Susino, L. Teriaca, and M. Uslenghi. Galactic and solar energetic particle observations during the increasing part of solar cycle 25 with EPD/HET and Metis on board Solar Orbiter. submitted to A&A, July 2025.

[6] Thomas K. Gaisser. *Cosmic Rays and Particle Physics*. 1991.

[7] C. Consolandi and AMS Collaboration. AMS-02 Monthly Proton Flux: Solar Modulation Effect and Short Time Scale Phenomena. In *34th International Cosmic Ray Conference (ICRC2015)*, volume 34 of *International Cosmic Ray Conference*, page 117, July 2015.

[8] M. Martucci, R. Munini, M. Boezio, V. Di Felice, O. Adriani, G. C. Barbarino, G. A. Bazilevskaya, R. Bellotti, M. Bongi, V. Bonvicini, S. Bottai, A. Bruno, F. Cafagna, D. Campana, P. Carlson, M. Casolino, G. Castellini, C. De Santis, A. M. Galper, A. V. Karelin, S. V. Koldashov, S. Koldobskiy, S. Y. Krutkov, A. N. Kvashnin, A. Leonov, V. Malakhov, L. Marcelli, N. Marcelli, A. G. Mayorov, W. Menn, M. Mergè, V. V. Mikhailov, E. Mocchiutti, A. Monaco, N. Mori, G. Osteria, B. Panico, P. Papini, M. Pearce, P. Picozza, M. Ricci, S. B. Ricciarini, M. Simon, R. Sparvoli, P. Spillantini, Y. I. Stozhkov, A. Vacchi, E. Vannuccini, G. Vasilyev, S. A. Voronov, Y. T. Yurkin, G. Zampa, N. Zampa, M. S. Potgieter, and J. L. Raath. Proton Fluxes Measured by the PAMELA Experiment from the Minimum to the Maximum Solar Activity for Solar Cycle 24. *Astrophysical Journal, Letters*, 854(1):L2, February 2018.

[9] M. Armano, H. Audley, J. Baird, M. Bassan, S. Benella, P. Binetruy, M. Born, D. Bortoluzzi, A. Cavalleri, A. Cesarini, A. M. Cruise, K. Danzmann, M. de Deus Silva, I. Diepholz, G. Dixon, R. Dolesi, M. Fabi, L. Ferraioli, V. Ferroni, N. Finetti, E. D. Fitzsimons, M. Freschi, L. Gesa, F. Gibert, D. Giardini, R. Giusteri, C. Grimani, J. Grzymisch, I. Harrison, G. Heinzel, M. Hewitson, D. Hollington, D. Hoyland, M. Hueller, H. Inchauspé, O. Jennrich, P. Jetzer, N. Karnesis, B. Kaune, N. Korsakova, C. J. Killow, M. Laurenza, J. A. Lobo, I. Lloro, L. Liu,

J. P. López-Zaragoza, R. Maarschalkerweerd, D. Mance, V. Martín, L. Martin-Polo, J. Martino, F. Martin-Porqueras, I. Mateos, P. W. McNamara, J. Mendes, L. Mendes, M. Nofrarias, S. Paczkowski, M. Perreur-Lloyd, A. Petiteau, P. Pivato, E. Plagnol, J. Ramos-Castro, J. Reiche, D. I. Robertson, F. Rivas, G. Russano, F. Sabbatini, J. Slutsky, C. F. Sopuerta, T. Sumner, D. Telloni, D. Texier, J. I. Thorpe, D. Vetrugno, S. Vitale, G. Wanner, H. Ward, P. Wass, W. J. Weber, L. Wissel, A. Wittchen, A. Zambotti, C. Zenoni, and P. Zweifel. Characteristics and Energy Dependence of Recurrent Galactic Cosmic-Ray Flux Depressions and of a Forbush Decrease with LISA Pathfinder. *Astrophysical Journal*, 854:113, February 2018.

[10] M. Armano, H. Audley, J. Baird, S. Benella, P. Binetruy, M. Born, D. Bortoluzzi, E. Castelli, A. Cavalleri, A. Cesarini, A. M. Cruise, K. Danzmann, M. de Deus Silva, I. Diepholz, G. Dixon, R. Dolesi, M. Fabi, L. Ferraioli, V. Ferroni, N. Finetti, E. D. Fitzsimons, M. Freschi, L. Gesa, F. Gibert, D. Giardini, R. Giusteri, C. Grimani, J. Grzymisch, I. Harrison, G. Heinzel, M. Hewitson, D. Hollington, D. Hoyland, M. Hueller, H. Inchauspé, O. Jennrich, P. Jetzer, N. Karnesis, B. Kaune, N. Korsakova, C. J. Killow, K. Kudela, M. Laurenza, J. A. Lobo, I. Lloro, L. Liu, J. P. López-Zaragoza, R. Maarschalkerweerd, D. Mance, N. Meshksar, V. Martín, L. Martin-Polo, J. Martino, F. Martin-Porqueras, I. Mateos, P. W. McNamara, J. Mendes, L. Mendes, M. Nofrarias, S. Paczkowski, M. Perreur-Lloyd, A. Petiteau, P. Pivato, E. Plagnol, J. Ramos-Castro, J. Reiche, D. I. Robertson, F. Rivas, G. Russano, J. Slutsky, C. F. Sopuerta, T. Sumner, D. Telloni, D. Texier, J. I. Thorpe, D. Vetrugno, M. Villani, S. Vitale, G. Wanner, H. Ward, P. Wass, W. J. Weber, L. Wissel, A. Wittchen, and P. Zweifel. Forbush Decreases and <2 Day GCR Flux Non-recurrent Variations Studied with LISA Pathfinder. *Astrophysical Journal*, 874(2):167, Apr 2019.

[11] Catia Grimani, Andrea Cesarini, Michele Fabi, Federico Sabbatini, Daniele Telloni, and Mattia Villani. Recurrent galactic cosmic-ray flux modulation in l1 and geomagnetic activity during the declining phase of the solar cycle 24. *Astrophysical Journal*, 904(1):64, nov 2020.

[12] Cheng-Rui Zhu and Mei-Juan Wang. Solar modulation of ams-02 daily proton and helium fluxes with modified force-field approximation models. *The Astrophysical Journal*, 980(1):116, feb 2025.

[13] C. Grimani, V. Andretta, P. Chioetto, V. Da Deppo, M. Fabi, S. Gissot, G. Naletto, A. Persici, C. Plainaki, M. Romoli, F. Sabbatini, D. Spadaro, M. Stangalini, D. Telloni, M. Uslenghi, E. Antonucci, A. Bemporad, G. Capobianco, G. Capuano, M. Casti, Y. De Leo, S. Fineschi, F. Frassati, F. Frassetto, P. Heinzel, G. Jerse, F. Landini, A. Liberatore, E. Magli, M. Messerotti, D. Moses, G. Nicolini, M. Pancrazzi, M. G. Pelizzo, P. Romano, C. Sasso, U. Schühle, A. Slemer, T. Straus, R. Susino, L. Teriaca, C. A. Volpicelli, J. L. Freiherr von Forstner, and P. Zuppella.

Cosmic-ray flux predictions and observations for and with Metis on board Solar Orbiter. *Astronomy and Astrophysics*, 656:A15, December 2021.

[14] C. Grimani, V. Andretta, E. Antonucci, P. Chioetto, V. Da Deppo, M. Fabi, S. Gissot, G. Jerse, M. Messerotti, G. Naletto, M. Pancrazzi, A. Persici, C. Plainaki, M. Romoli, F. Sabbatini, D. Spadaro, M. Stangalini, D. Telloni, L. Teriaca, M. Uslenghi, M. Villani, L. Abbo, A. Burtovoi, F. Frassati, F. Landini, G. Nicolini, G. Russano, C. Sasso, and R. Susino. Particle monitoring capability of the Solar Orbiter Metis coronagraph through the increasing phase of solar cycle 25. *Astronomy and Astrophysics*, 677:A45, September 2023.

[15] C. Grimani, M. Fabi, A. Persici, F. Sabbatini, M. Villani, F. Frassati, E. Antonucci, M. Pancrazzi, D. Telloni, P. Kühl, J. Rodríguez-Pacheco, R. F. Wimmer-Schweingruber, V. Andretta, P. Chioetto, V. Da Deppo, S. Gissot, G. Jerse, M. Messerotti, G. Naletto, C. Plainaki, M. Romoli, D. Spadaro, M. Stangalini, L. Teriaca, M. Uslenghi, L. Abbo, A. Burtovoi, F. Landini, G. Nicolini, G. Russano, C. Sasso, and R. Susino. Observation of solar energetic particles with Metis on board Solar Orbiter on February 25, 2023. *Astronomy and Astrophysics*, 686:A74, June 2024.

[16] C Grimani, M Fabi, F Sabbatini, and M Villani. Space weather predictions for LISA during solar cycle 26. *Classical and Quantum Gravity*, 42(9):095009, apr 2025.

[17] O. Adriani, G. C. Barbarino, G. A. Bazilevskaya, R. Bellotti, M. Boezio, E. A. Bogomolov, L. Bonechi, M. Bongi, V. Bonvicini, S. Borisov, S. Bottai, A. Bruno, F. Cafagna, D. Campana, R. Carbone, P. Carlson, M. Casolino, G. Castellini, L. Consiglio, M. P. De Pascale, C. De Santis, N. De Simone, V. Di Felice, V. Formato, A. M. Galper, L. Grishantseva, W. Gillard, G. Jerse, A. V. Karelin, S. V. Koldashov, S. Y. Krutkov, A. N. Kvashnin, A. Leonov, V. Malakhov, L. Marcelli, A. G. Mayorov, W. Menn, V. V. Mikhailov, E. Mocchiutti, A. Monaco, N. Mori, N. Nikonov, G. Osteria, F. Palma, P. Papini, M. Pearce, P. Picozza, C. Pizzolotto, M. Ricci, S. B. Ricciarini, R. Sarkar, L. Rossetto, M. Simon, R. Sparvoli, P. Spillantini, Y. I. Stozhkov, A. Vacchi, E. Vannuccini, G. Vasilyev, S. A. Voronov, J. Wu, Y. T. Yurkin, G. Zampa, N. Zampa, and V. G. Zverev. Observations of the 2006 December 13 and 14 Solar Particle Events in the 80 MeV $n^{-1}$-3 GeV $n^{-1}$ Range from Space with the PAMELA Detector. *Astrophysical Journal*, 742(2):102, December 2011.

[18] Donald V. Reames. *Solar Energetic Particles. A Modern Primer on Understanding Sources, Acceleration and Propagation*, volume 978. 2021.

[19] A. Kumari, D. E. Morosan, E. K. J. Kilpua, and F. Daei. Type II radio bursts and their association with coronal mass ejections in solar cycles 23 and 24. *Astronomy and Astrophysics*, 675:A102, July 2023.

[20] R.A. Nymmik. SEP Event Distribution Function as Inferred from Spaceborne

Measurements and Lunar Rock Isotopic Data. In *26th Int. Cosmic Ray Conf. (Salt Lake City)*, volume 6, pages 268–271, 1999a.

[21] R.A. Nymmik. Relationships among Solar Activity, SEP Occurrence Frequency, and Solar Energetic Particle Event Distribution Function. In *26th Int. Cosmic Ray Conf. (Salt Lake City)*, volume 6, pages 280–283, 1999b.

[22] Catia Grimani, Michele Fabi, Federico Sabbatini, Mattia Villani, Luca Antognini, Aishah Bashiri, Lucio Calcagnile, Anna Paola Caricato, Roberto Catalano, Deborah Chilà, Giuseppe Antonio Pablo Cirrone, Tommaso Croci, Giacomo Cuttone, Sylvain Dunand, Luca Frontini, Maria Ionica, Keida Kanxheri, Matthew Large, Valentino Liberali, Maurizio Martino, Giuseppe Maruccio, Giovanni Mazza, Mauro Menichelli, Anna Grazia Monteduro, Arianna Morozzi, Francesco Moscatelli, Stefania Pallotta, Daniele Passeri, Maddalena Pedio, Marco Petasecca, Giada Petringa, Francesca Peverini, Lorenzo Piccolo, Pisana Placidi, Gianluca Quarta, Silvia Rizzato, Alberto Stabile, Cinzia Talamonti, Jonathan Thomet, Luca Tosti, Richard James Wheadon, Nicolas Wyrsch, Nicola Zema, and Leonello Servoli. A hydrogenated amorphous silicon detector for Space Weather applications. *Astrophysics and Space Science*, 368(9):78, September 2023.

[23] A. Papaioannou, A. Kouloumvakos, A. Mishev, R. Vainio, I. Usoskin, K. Herbst, A. P. Rouillard, A. Anastasiadis, J. Gieseler, R. Wimmer-Schweingruber, and P. Kühl. The first ground-level enhancement of solar cycle 25 on 28 October 2021. *Astronomy and Astrophysics*, 660:L5, April 2022.

[24] Francesco Faldi. *Study of short term variations in the cosmic proton fluxes with the AMS–02 experiment.* PhD thesis, University of Perugia, 2025.

[25] Francesco Faldi, Bruna Bertucci, Nicola Tomassetti, and Valerio Vagelli. Real-time monitoring of solar energetic particles outside the ISS with the AMS-02 instrument. *Rendiconti Lincei. Scienze Fisiche e Naturali*, 34(2):339–345, June 2023.

[26] Joseph O'Neill, Fan Lei, Keith Ryden, Paul Morris, Ben Clewer, Fraser Baird, Paul Sellin, Clive Dyer, Melanie Heil, Piers Jiggens, and Giovanni Santin. The high-energy proton instrument (HEPI), a compact Cherenkov radiation space weather monitor. *Journal of Space Weather and Space Climate*, 15:27, January 2025.

[27] Pau Amaro-Seoane, Heather Audley, Stanislav Babak, John Baker, Enrico Barausse, Peter Bender, Emanuele Berti, Pierre Binetruy, Michael Born, Daniele Bortoluzzi, Jordan Camp, Chiara Caprini, Vitor Cardoso, Monica Colpi, John Conklin, Neil Cornish, Curt Cutler, Karsten Danzmann, Rita Dolesi, Luigi Ferraioli, Valerio Ferroni, Ewan Fitzsimons, Jonathan Gair, Lluis Gesa Bote, Domenico Giardini, Ferran Gibert, Catia Grimani, Hubert Halloin, Gerhard Heinzel, Thomas Hertog, Martin Hewitson, Kelly Holley-Bockelmann, Daniel Hollington, Mauro Hueller, Henri Inchauspe, Philippe Jetzer, Nikos Karnesis, Christian Killow, Antoine Klein, Bill Klipstein, Natalia Korsakova, Shane L Larson, Jeffrey Livas, Ivan Lloro, Nary Man, Davor Mance, Joseph Martino, Ignacio Mateos, Kirk McKenzie, Sean T McWilliams, Cole Miller, Guido Mueller, Germano

Nardini, Gijs Nelemans, Miquel Nofrarias, Antoine Petiteau, Paolo Pivato, Eric Plagnol, Ed Porter, Jens Reiche, David Robertson, Norna Robertson, Elena Rossi, Giuliana Russano, Bernard Schutz, Alberto Sesana, David Shoemaker, Jacob Slutsky, Carlos F. Sopuerta, Tim Sumner, Nicola Tamanini, Ira Thorpe, Michael Troebs, Michele Vallisneri, Alberto Vecchio, Daniele Vetrugno, Stefano Vitale, Marta Volonteri, Gudrun Wanner, Harry Ward, Peter Wass, William Weber, John Ziemer, and Peter Zweifel. Laser Interferometer Space Antenna. *arXiv e-prints*, page arXiv:1702.00786, Feb 2017.

[28] Monica Colpi, Karsten Danzmann, Martin Hewitson, Kelly Holley-Bockelmann, Philippe Jetzer, Gijs Nelemans, Antoine Petiteau, David Shoemaker, Carlos Sopuerta, Robin Stebbins, et al. LISA: Definition study report. *arXiv e-prints*, 2024.

[29] David Mazzanti, Daniel Guberman, Angels Aran, Lluis Garrido, David Gascon, Federico Izraelevitch, Joan Mauricio, David Roma, Víctor Martín, and Miquel Nofrarias. A low-power SiPM-based radiation monitor for LISA. *PoS*, ICRC2023:1494, 2023.

[30] E. Garutti and Yu. Musienko. Radiation damage of SiPMs. *Nuclear Instruments and Methods in Physics Research A*, 926:69–84, May 2019.

[31] Xutao Zheng, Huaizhong Gao, Jiaxing Wen, Ming Zeng, Xiaofan Pan, Dacheng Xu, Yihui Liu, Yuchong Zhang, Haowei Peng, Yuchen Jiang, Xiangyun Long, Di'an Lu, Dongxin Yang, Hua Feng, Zhi Zeng, Jirong Cang, and Yang Tian. In-orbit radiation damage characterization of sipms in the grid-02 cubesat detector. *Nuclear Instruments and Methods in Physics Research Section A: Accelerators, Spectrometers, Detectors and Associated Equipment*, 1044:167510, 2022.

[32] Catia Grimani, Mattia Villani, Michele Fabi, and Federico Sabbatini. LISA and LISA-like mission test-mass charging for gamma-ray burst detection. *Journal of High Energy Astrophysics*, 42:38–51, June 2024.

[33] J.R. Srour, G.J. Vendura, D.H. Lo, C.M. Chantal Toporow. Ph.D., M. Dooley, R.P. Nakano, and E.E. King. Damage mechanisms in radiation-tolerant amorphous silicon solar cells. *Nuclear Science, IEEE Transactions on*, 45:2624 – 2631, 01 1999.

[34] M. Menichelli, M. Bizzarri, M. Boscardin, L. Calcagnile, M. Caprai, A.P. Caricato, G.A.P. Cirrone, M. Crivellari, I. Cupparo, G. Cuttone, S. Dunand, L. Fanò, B. Gianfelici, O. Hammad, M. Movileanu-Ionica, M. Large, G. Maruccio, A.G. Monteduro, F. Moscatelli, A. Morozzi, A. Papi, D. Passeri, M. Pedio, M. Petasecca, G. Petringa, F. Peverini, G. Quarta, S. Rizzato, A. Rossi, A. Scorzoni, L. Servoli, C. Talamonti, G. Verzellesi, and Wyrsch N. Displacement damage in Hydrogenated Amorphous Silicon p-i-n diodes and charge selective contacts detectors. *TechRxiv*, July 2022.

[35] Nicolas Wyrsch, C. Miazza, S. Dunand, Christophe Ballif, Arvind Shah, M. Despeisse, D. Moraes, F. Powolny, and P. Jarron. Radiation hardness of

amorphous silicon particle sensors. *Journal of Non-Crystalline Solids*, 352:1797–1800, 06 2006.

[36] M. Menichelli, M. Bizzarri, L. Calcagnile, M. Caprai, A. P. Caricato, R. Catalano, G. A. P. Cirrone, T. Croci, G. Cuttone, S. Dunand, M. Fabi, L. Frontini, B. Gianfelici, C. Grimani, M. Ionica, K. Kanxheri, M. Large, V. Liberali, M. Martino, G. Maruccio, G. Mazza, A. G. Monteduro, A. Morozzi, F. Moscatelli, S. Pallotta, A. Papi, D. Passeri, M. Pedio, M. Petasecca, G. Petringa, F. Peverini, L. Piccolo, P. Placidi, G. Quarta, S. Rizzato, G. Rossi, F. Sabbatini, A. Stabile, L. Servoli, C. Talamonti, M. Villani, R. J. Wheadon, and N. Wyrsch. Development of thin hydrogenated amorphous silicon detectors on a flexible substrate. *arXiv e-prints*, page arXiv:2211.17114, November 2022.

[37] M. Menichelli, L. Antognini, A. Bashiri, M. Bizzarri, L. Calcagnile, M. Caprai, A.P. Caricato, R. Catalano, G.A.P. Cirrone, T. Croci, G. Cuttone, S. Dunand, M. Fabi, L. Frontini, C. Grimani, M. Ionica, K. Kanxheri, M. Large, V. Liberali, M. Martino, G. Maruccio, G. Mazza, A. G. Monteduro, A. Morozzi, F. Moscatelli, S. Pallotta, A. Papi, D. Passeri, M. Pedio, M. Petasecca, G. Petringa, F. Peverini, L. Piccolo, P. Placidi, G. Quarta, S. Rizzato, G. Rossi, F. Sabbatini, A. Stabile, L. Servoli, C. Talamonti, L. Tosti, M. Villani, R.J. Wheadon, N. Wyrsch, and N. Zema. X-ray qualification of hydrogenated amorphous silicon sensors on flexible substrate. In *2023 9th International Workshop on Advances in Sensors and Interfaces (IWASI)*, pages 190–193, 2023.

[38] Mauro Menichelli, Luca Antognini, Saba Aziz, Aishah Bashiri, Marco Bizzarri, Lucio Calcagnile, Mirco Caprai, Domenico Caputo, Anna Paola Caricato, Roberto Catalano, Deborah Chilà, Giuseppe Antonio Pablo Cirrone, Tommaso Croci, Giacomo Cuttone, Giampiero De Cesare, Sylvain Dunand, Michele Fabi, Luca Frontini, Catia Grimani, Maria Ionica, Keida Kanxheri, Matthew Large, Valentino Liberali, Nicola Lovecchio, Maurizio Martino, Giuseppe Maruccio, Giovanni Mazza, Anna Grazia Monteduro, Arianna Morozzi, Francesco Moscatelli, Augusto Nascetti, Stefania Pallotta, Andrea Papi, Daniele Passeri, Maddalena Pedio, Marco Petasecca, Giada Petringa, Francesca Peverini, Lorenzo Piccolo, Pisana Placidi, Gianluca Quarta, Silvia Rizzato, Giulia Rossi, Federico Sabbatini, Leonello Servoli, Alberto Stabile, Cinzia Talamonti, Jonathan Emanuel Thomet, Luca Tosti, Mattia Villani, Richard J. Wheadon, Nicolas Wyrsch, and Nicola Zema. Characterization of hydrogenated amorphous silicon sensors on polyimide flexible substrate. *IEEE Sensors Journal*, 24(8):12466–12471, 2024.

[39] Matthew James Large, Keida Kanxheri, Jessie Posar, Saba Aziz, Aishah Bashiri, Lucio Calcagnile, Daniela Calvo, Domenico Caputo, Anna Paola Caricato, Roberto Catalano, Roberto Cirio, Giuseppe Antonio Pablo Cirrone, Tommaso Croci, Giacomo Cuttone, Gianpiero De Cesare, Paolo De Remigis, Sylvain Dunand, Michele Fabi, Luca Frontini, Catia Grimani, Mariacristina Guarrera, Maria Ionica, Francesca Lenta, Valentino Liberali, Nicola Lovecchio, Maurizio Martino, Giuseppe

Maruccio, Giovanni Mazza, Mauro Menichelli, Anna Grazia Monteduro, Arianna Morozzi, Francesco Moscatelli, Augusto Nascetti, Stefania Pallotta, Daniele Passeri, Maddalena Pedio, Giada Petringa, Francesca Peverini, Pisana Placidi, Gianluca Quarta, Silvia Rizzato, Federico Sabbatini, Leonello Servoli, Alberto Stabile, Jonathan Emanuel Thomet, Luca Tosti, Mattia Villani, Richard James Wheadon, Nicolas Wyrsch, Nicola Zema, Marco Petasecca, and Cinzia Talamonti. Dosimetry of microbeam radiotherapy by flexible hydrogenated amorphous silicon detectors. *Physics in Medicine Biology*, 69(15):155022, jul 2024.

[40] Vasilis Vlachoudis. Flair: A powerful but user friendly graphical interface for FLUKA. In *International Conference on Mathematics, Computational Methods & Reactor Physics (M&C 2009), Saratoga Springs, New York*, pages 790–800, 2009.

[41] Giuseppe Battistoni, Till Boehlen, Francesco Cerutti, Pik Wai Chin, Luigi Salvatore Esposito, Alberto Fassò, Alfredo Ferrari, Alessio Mereghetti, Pablo Garcia Ortega, Johannes Ranft, Stefan Roesler, Paola R. Sala, and Vasilis Vlachoudis. Overview of the FLUKA code. In *Joint International Conference on Supercomputing in Nuclear Applications + Monte Carlo*, page 06005, June 2014.

[42] T. T. Böhlen, F. Cerutti, M. P. W. Chin, A. Fassò, A. Ferrari, P. G. Ortega, A. Mairani, P. R. Sala, G. Smirnov, and V. Vlachoudis. The FLUKA Code: Developments and Challenges for High Energy and Medical Applications. *Nuclear Data Sheets*, 120:211–214, June 2014.

[43] P. Thirupathaiah, Siddhi Y. Shah, and S.A. Haider. Characteristics of solar x-ray flares and their effects on the ionosphere and human exploration to mars: Mgs radio science observations. *Icarus*, 330:60–74, 2019.

[44] Bernard Gottschalk. Charge-balancing current integrator with large dynamic range. *Nuclear Instruments and Methods in Physics Research*, 207(3):417–421, April 1983.

[45] G. Mazza, R. Cirio, M. Donetti, A. La Rosa, A. Luparia, F. Marchetto, and C. Peroni. A 64-Channel Wide Dynamic Range Charge Measurement ASIC for Strip and Pixel Ionization Detectors. *IEEE Transactions on Nuclear Science*, 52(4):847–853, August 2005.

[46] F. Dimiccoli, R. Dolesi, M. Fabi, V. Ferroni, C. Grimani, M. Muratore, P. Sarra, M. Villani, and W. J. Weber. LISA test-mass charging: Particle flux modeling, Monte Carlo simulations, and induced effects on the sensitivity of the observatory. *Astronomy and Astrophysics*, 700:A102, August 2025.

[47] C. Grimani, M. Fabi, N. Finetti, D. Tombolato, L. Marconi, et al. Heliospheric influences on LISA. *Classical and Quantum Gravity.*, 26:094018, 2009.

[48] J. Rodríguez-Pacheco, R. F. Wimmer-Schweingruber, G. M. Mason, G. C. Ho, Sánchez-Prieto, S., Prieto, M., Martín, C., Seifert, H., Andrews, G. B., Kulkarni, S. R., Panitzsch, L., Boden, S., Böttcher, S. I., Cernuda, I., Elftmann, R., Espinosa Lara, F., Gómez-Herrero, R., Terasa, C., Almena, J., Begley, S., Böhm, E., Blanco, J. J., Boogaerts, W., Carrasco, A., Castillo, R., da Silva Fariña, A., de Manuel

González, V., Drews, C., Dupont, A. R., Eldrum, S., Gordillo, C., Gutiérrez, O., Haggerty, D. K., Hayes, J. R., Heber, B., Hill, M. E., Jüngling, M., Kerem, S., Knierim, V., Köhler, J., Kolbe, S., Kulemzin, A., Lario, D., Lees, W. J., Liang, S., Martínez Hellín, A., Meziat, D., Montalvo, A., Nelson, K. S., Parra, P., Paspirgilis, R., Ravanbakhsh, A., Richards, M., Rodríguez-Polo, O., Russu, A., Sánchez, I., Schlemm, C. E., Schuster, B., Seimetz, L., Steinhagen, J., Tammen, J., Tyagi, K., Varela, T., Yedla, M., Yu, J., Agueda, N., Aran, A., Horbury, T. S., Klecker, B., Klein, K.-L., Kontar, E., Krucker, S., Maksimovic, M., Malandraki, O., Owen, C. J., Pacheco, D., Sanahuja, B., Vainio, R., Connell, J. J., Dalla, S., Dröge, W., Gevin, O., Gopalswamy, N., Kartavykh, Y. Y., Kudela, K., Limousin, O., Makela, P., Mann, G., Önel, H., Posner, A., Ryan, J. M., Soucek, J., Hofmeister, S., Vilmer, N., Walsh, A. P., Wang, L., Wiedenbeck, M. E., Wirth, K., and Zong, Q. The Energetic Particle Detector - Energetic particle instrument suite for the Solar Orbiter mission. *Astronomy and Astrophysics*, 642:A7, 2020.

[49] C. Grimani, X. Ao, M. Fabi, M. Laurenza, G. Li, A. Lobo, I. Mateos, M. Storini, O. Verkhoglyadova, and G. P. Zank. LISA-PF radiation monitor performance during the evolution of SEP events for the monitoring of test-mass charging. *Classical and Quantum Gravity*, 31(4):045018, February 2014.

[50] Jan Gieseler, Nina Dresing, Christian Palmroos, Johan L. Freiherr von Forstner, Daniel J. Price, Rami Vainio, Athanasios Kouloumvakos, Laura Rodríguez-García, Domenico Trotta, Vincent Génot, Arnaud Masson, Markus Roth, and Astrid Veronig. Solar-MACH: An open-source tool to analyze solar magnetic connection configurations. *Frontiers in Astronomy and Space Sciences*, 9, 2023.

[51] Krucker, Säm, Hurford, G. J., Grimm, O., Kögl, S., Gröbelbauer, H.-P, Etesi, L., Casadei, D., Csillaghy, A., Benz, A. O., Arnold, N. G., Molendini, F., Orleanski, P., Schori, D., Xiao, H., Kuhar, M., Hochmuth, N., Felix, S., Schramka, F., Marcin, S., Kobler, S., Iseli, L., Dreier, M., Wiehl, H. J., Kleint, L., Battaglia, M., Lastufka, E., Sathiapal, H., Lapadula, K., Bednarzik, M., Birrer, G., Stutz, St., Wild, Ch., Marone, F., Skup, K. R., Cichocki, A., Ber, K., Rutkowski, K., Bujwan, W., Juchnikowski, G., Winkler, M., Darmetko, M., Michalska, M., Seweryn, K., Bialek, A., Osica, P., Sylwester, J., Kowalinski, M., ´Scislowski, D., Siarkowski, M., Ste´slicki, M., Mrozek, T., Podgórski, P., Meuris, A., Limousin, O., Gevin, O., Le Mer, I., Brun, S., Strugarek, A., Vilmer, N., Musset, S., Maksimovi´c, M., Fárník, F., Kozácek, Z., Kasparová, J., Mann, G., Önel, H., Warmuth, A., Rendtel, J., Anderson, J., Bauer, S., Dionies, F., Paschke, J., Plüschke, D., Woche, M., Schuller, F., Veronig, A. M., Dickson, E. C. M., Gallagher, P. T., Maloney, S. A., Bloomfield, D. S., Piana, M., Massone, A. M., Benvenuto, F., Massa, P., Schwartz, R. A., Dennis, B. R., van Beek, H. F., Rodríguez-Pacheco, J., and Lin, R. P. The Spectrometer/Telescope for Imaging X-rays (STIX). *Astronomy and Astrophysics*, 642:A15, 2020.